\documentclass[twocolumn]{aastex63}
\def \trg       {PKS\,1830}
\def \artip     {\texttt{ARTIP}}

\def \kms       {km~s$^{-1}$}
\def \cmc       {cm$^{-3}$}
\def \emuni     {cm$^{-6}$~pc}
\def \deg       {\text{$^{\circ}$}}

\def \asec      {\text{$^{\prime\prime}$}}

\def \mjyb      {mJy~beam$^{-1}$}
\def \Hii       {\text{H~\textsc{ii}}}

\def \Hi        {\text{H \textsc{i}}}
\def \Lsun      {\text{$\rm L_{\odot}$}}
\def \Msun      {\text{$\rm M_{\odot}$}}
\def \Msunpc    {\text{$\rm M_{\odot} \, pc^{-2}$}}
\def \Msunyr    {\text{$\rm M_{\odot} \, yr^{-1}$}}
\def \QAuni     {\text{$\rm photons \, s^{-1} \, pc^{-2}$}}

\received{...}
\revised{...}
\accepted{...}
\submitjournal{ApJ}
\shorttitle{Radio Recombination Lines at $z=0.89$}
\shortauthors{Emig et al.}

\begin{document}

\title{Discovery of Hydrogen Radio Recombination Lines at $z=0.89$ towards PKS\, 1830$-$211}

%% Use \affiliation for affiliation information. The old \affil is now aliased
%% to \affiliation. AASTeX v6.3 will automatically index these in the header.
%% When a duplicate is found its index will be the same as its previous entry.
%%
%% The new \altaffiliation can be used to indicate some secondary information
%% such as fellowships. This command produces a non-numeric footnote that is
%% set away from the numeric \affiliation footnotes.  NOTE that if an
%% \altaffiliation command is used it must come BEFORE the \affiliation call,
%% right after the \author command, in order to place the footnotes in
%% the proper location.
%%

\correspondingauthor{Kimberly L. Emig}
\email{kemig@nrao.edu}

\author[0000-0001-6527-6954]{Kimberly L. Emig}
\altaffiliation{Jansky Fellow of the National Radio Astronomy Observatory}
\affiliation{National Radio Astronomy Observatory, 520 Edgemont Road, Charlottesville, VA 22903, USA}

\author[0000-0001-7547-4241]{Neeraj Gupta}
\affiliation{Inter-University Centre for Astronomy and Astrophysics, Post Bag 4, Ganeshkhind, Pune 411 007, India}

\author{Pedro Salas}
\affiliation{Green Bank Observatory, 155 Observatory Road, Green Bank, WV 24915, USA}

\author{S\'ebastien Muller}
\affiliation{Department of Space, Earth and Environment, Chalmers University of Technology, Onsala Space Observatory, SE-43992 Onsala, Sweden}

\author{Sergei A.~Balashev}
\affiliation{Ioffe Institute, Politekhnicheskaya 26, 194021 Saint Petersburg, Russia}
\affiliation{HSE University, Saint Petersburg, 190121, Russia}

\author{Fran\c{c}oise Combes}
\affiliation{Observatoire de Paris, Coll\`ege de France, PSL University, Sorbonne University, CNRS, LERMA, Paris, France}

\author{Emmanuel Momjian}
\affiliation{National Radio Astronomy Observatory, 1003 Lopezville Road, Socorro, NM 87801, USA}

\author{Yiqing Song}
\affiliation{Department of Astronomy, University of Virginia, 530 McCormick Road, Charlottesville, VA 22903, USA}
\affiliation{National Radio Astronomy Observatory, 520 Edgemont Road, Charlottesville, VA 22903, USA}

\author{Preshanth Jagannathan}
\affiliation{National Radio Astronomy Observatory, 1003 Lopezville Road, Socorro, NM 87801, USA}

\author[0000-0001-9174-1186]{Partha P.~Deka}
\affiliation{Inter-University Centre for Astronomy and Astrophysics, Post Bag 4, Ganeshkhind, Pune 411 007, India}

\author{Gyula I.~G.~J\'ozsa}
\affiliation{Max-Planck Institut f\"{u}r Radioastronomie, Auf dem H\"{u}gel 69, 53121 Bonn, Germany}
\affiliation{Department of Physics and Electronics, Rhodes University, PO Box 94, Makhanda, 6140, South Africa}

\author{Hans-Rainer Kl\"{o}ckner}
\affiliation{Max-Planck Institut f\"{u}r Radioastronomie, Auf dem H\"{u}gel 69, 53121 Bonn, Germany}

\author{Abhisek Mohapatra}
\affiliation{Inter-University Centre for Astronomy and Astrophysics, Post Bag 4, Ganeshkhind, Pune 411 007, India}

\author{Pasquier Noterdaeme}
\affiliation{Institut d’Astrophysique de Paris, Sorbonne Universit\'e and CNRS, 98bis boulevard Arago, F-75014 Paris, France}
\affiliation{Franco-Chilean Laboratory for Astronomy, IRL 3386, CNRS and U. de Chile, Casilla 36-D, Santiago, Chile}

\author{Patrick Petitjean}
\affiliation{Institut d’Astrophysique de Paris, Sorbonne Universit\'e and CNRS, 98bis boulevard Arago, F-75014 Paris, France}

\author{Raghunathan Srianand}
\affiliation{Inter-University Centre for Astronomy and Astrophysics, Post Bag 4, Ganeshkhind, Pune 411 007, India}

\author{Jonah D.~Wagenveld}
\affiliation{Max-Planck Institut f\"{u}r Radioastronomie, Auf dem H\"{u}gel 69, 53121 Bonn, Germany}

%%%%%%%%%%%%%%%%%%%%%%%%%%%%%%%%%%%%%%%%

\begin{abstract}

We report the detection of stimulated hydrogen radio recombination line (RRL) emission from ionized gas in a $z=0.89$ galaxy using 580--1670 MHz observations from the MeerKAT Absorption Line Survey (MALS). The RRL emission originates in a galaxy that intercepts and strongly lenses the radio blazar \trg{}$-$211 ($z=2.5$). This is the second detection of RRLs outside of the local universe and the first clearly associated with hydrogen. We detect effective H144$\alpha$ (and H163$\alpha$) transitions at observed frequencies of 1156~(798)~MHz by stacking 17~(27) RRLs with 21$\sigma$ (14$\sigma$) significance. The RRL emission contains two main velocity components and is coincident in velocity with \Hi{} 21~cm and OH 18~cm absorption. We use the RRL spectral line energy distribution and a Bayesian analysis to constrain the density ($n_e$) and the volume-averaged pathlength ($\ell$) of the ionized gas. We determine $\log( n_e ) = 2.0_{-0.7}^{+1.0}$~\cmc{} and $\log( \ell ) = -0.7_{-1.1}^{+1.1}$~pc towards the north east (NE) lensed image, likely tracing the diffuse thermal phase of the ionized ISM in a thin disk. Towards the south west (SW) lensed image, we determine $\log( n_e ) = 3.2_{-1.0}^{+0.4}$~\cmc{} and $\log( \ell ) = -2.7_{-0.2}^{+1.8}$~pc, tracing gas that is more reminiscent of \Hii{} regions.
We estimate a star formation (surface density) rate of $\Sigma_{\rm SFR} \sim 0.6$~\Msunyr{}~kpc$^{-2}$ or SFR~$\sim 50$~\Msunyr{}, consistent with a star-forming main sequence galaxy of  $M_{\star} \sim 10^{11}$~\Msun{}. 
The discovery presented here opens up the possibility of studying ionized gas at high redshifts using RRL observations from current and future (e.g., SKA and ngVLA) radio facilities.

\end{abstract}


%%%%%%%%%%%%%%%%%%%%%%%%%%%%%%%%%%%%%%%%

\section{Introduction} \label{sec:intro}

Radio recombination lines (RRLs) result from the radiative de-excitation of electrons at high excitation levels of atoms. RRLs with frequencies $\nu_{\mathrm{rest}} \lesssim $10~GHz in extragalactic sources probe gas with relatively low density that can be stimulated by radio continuum \citep[for an overview, see][]{Emig2021}, upon which the line emission is significantly enhanced compared with the local thermodynamic equilibrium (LTE) Boltzmann distribution. The physical conditions of the gas strongly influence which principal quantum numbers (thus frequencies) show enhanced emission. Accordingly, the relative intensities of RRLs, the so-called spectral line energy distribution (SLED), carry information on the temperature, density, and pathlength of the diffuse gas component \citep[e.g.,][]{Shaver1975a, Salgado2017a, Oonk2017}. Stimulated emission has the added benefit that its intensity is proportional to the background continuum intensity at a given frequency, $S_{\mathrm{RRL}} \propto S_{\mathrm{bkg \, cont}}$. Therefore, it is conceivable to observe these RRLs at cosmological distances wherever bright radio sources are present \citep{Shaver1978b}. In contrast, RRLs at high radio frequencies ($\gtrsim 10$~GHz) typically trace spontaneous recombination emission (in LTE), making them a great direct measure of ionizing photons and therefore star formation rates. However, the flux of spontaneous transitions falls off with distance ($D$) and frequency ($\nu$) as $S \propto D^{-2} \nu^{-1}$, limiting the observation of this faint emission to galaxies in the nearby universe.

The very first extragalactic detections of RRLs, in M82 and NGC253, found contributions from stimulation \citep{Shaver1977, Seaquist1977, Shaver1978a}. In the case of M82, \cite{Bell1978a} modeled the SLED of hydrogen RRLs and showed that the increasing intensity of the RRL emission at $\nu < 10$~GHz can only result due to stimulation by the large-scale synchrotron continuum of the galaxy. They determined that the hydrogen RRLs originate in ionized gas with $n_e \approx 150$~\cmc{} and pathlength $\ell \approx 110$~pc. Soon after, \cite{Churchwell1979} used the Arecibo 300~m telescope to search 21 galaxies and active galactic nuclei (AGN) for RRL emission with the 1.4 GHz receiver and three AGN with the 430 MHz receiver, with the set-up covering just a single RRL transition. They did not detect emission with line-to-continuum ratios of $\tau_{\mathrm{RRL}} > 10^{-3}$. To similar sensitivities, \citet{Bell1984} used the Effelsberg 100~m dish at 4.8~GHz to search ten galaxies without clear success. \citet{Bell1980} discovered the H83$\alpha$ and H99$\alpha$ lines at 10.5 GHz and 6.2 GHz, respectively, in the GHz peaked-spectrum source OQ 208 at $z=0.0763$, showing that this RRL emission could also only arise due to stimulation. These studies used narrow bandwidth receivers and were only sensitive to one RRL spectral line per observation.

To date, 8 of the 23 external galaxies with detected RRL emission show evidence for stimulated emission by non-thermal emission \citep[for an overview, see][]{Emig2021}\footnote{We also refer the reader for an update collection of extragalactic RRL detections at \url{www.tinyurl.com/DatabaseForExtragalacticRRLs}}. These were all local ($D < 350$~Mpc) and mostly star forming galaxies, until recently, \cite{Emig2019} used the Low Frequency Array to detect stimulated RRLs at $z = 1.124$ with a rest-frame frequency of 284 MHz. They argue that the RRL emission most likely arises from carbon in an intervening galaxy along the line-of-sight to 3C\,190. However, they could not clearly discern whether the emission was from carbon or hydrogen since the lines are separated by 150~\kms{} and have comparable intensities at those frequencies in the Milky Way \citep[e.g.,][]{Anantharamaiah1985a}. The improved sensitivity of high-resolution interferometers and the large fractional bandwidths that enable deeper searches through line stacking are making extragalactic RRL detections now feasible. Furthermore, given that the frequency separation between each RRL is unique, the development of new cross-correlation techniques are enabling blind searches of RRLs across redshift space \citep{Emig2020a}. 

\begin{figure}
    \centering
    \includegraphics[width=0.47\textwidth]{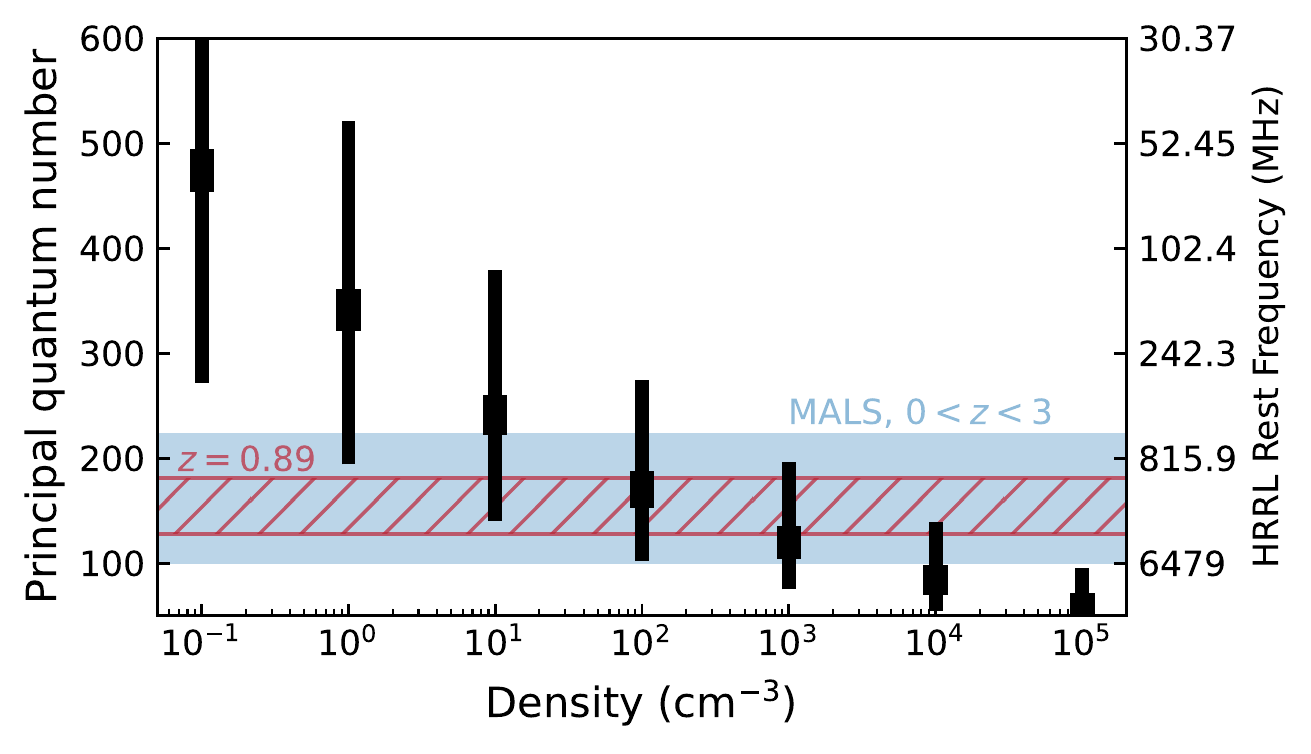}
    \caption{The principal quantum numbers, $\mathsf{n}$, of Hydrogen RRLs at which the stimulated-only emission of gas with a given density peaks. The thick black region marks the maximum $\mathsf{n}$ in the peak of emission, for ionized gas temperatures $ 5000~\mathrm{K} \leq T_e \leq 12\,000$~K, and the thin black region indicates the $\mathsf{n}$ for which the peak intensity is more than one half of the maximum. The shaded blue region shows the parameter space covered by MALS for RRL redshifts of $0 \leq z \leq 3$, while the red hatched region shows the coverage for a $z=0.89$ observation. To calculate the peak optical depths we do not take into account an external radiation field (which has a minimal effect), but do take into account collisional broadening of the line profile occurring in dense gas. }
    \label{fig:qn_ne}
\end{figure}

Current large-bandwidth spectral line surveys, such as the MeerKAT Absorption Line Survey \citep[MALS;][]{Gupta2016}, the First Large Absorption line Survey \citep[FLASH;][]{Allison2022}, the Search for HI Absorption with APERTIF \citep[SHARP; e.g., see ][]{Morganti2018} -- and in the future with the Square Kilometer Array \citep[SKA; e.g.,][]{Blyth2015} -- can observe tens of RRLs simultaneously, opening the way for ionized gas studies with optimum sensitivity to gas with electron densities of $1 ~\mathrm{cm^{-3}} \lesssim n_e \lesssim 10^4$~\cmc{} (e.g., see Fig.~\ref{fig:qn_ne}). These surveys can explore RRLs as a tracer of (diffuse) ionized gas in external galaxies for the first time in a large systematic way and address at what level stimulated RRLs are present in galaxies. Ultimately, these RRL observations will bring new insight into the evolution of the (ionized) interstellar medium of galaxies and the environment of AGN.

In particular, MALS is carrying out the most sensitive search to date ($\sigma \sim 0.6$~\mjyb{} per 6~\kms{} channel) of \Hi\ 21~cm and OH 18~cm absorption lines at $0 \lesssim z \lesssim 2$ \citep{Gupta2016}. MALS is observing $\sim$500 pointings centered on the brightest ($S_{\mathrm{1\, GHz}}>0.2$~Jy) radio sources at declination $\delta \lesssim +30^{\circ}$ \citep[see][for the survey footprint]{Gupta2022}, using the MeerKAT \citep{Jonas2016} L band, nominally covering 900--1670 MHz, and UHF band, nominally covering 580--1015 MHz. Towards each sight line, the survey is sensitive to peak RRL line to continuum ratios of $\tau_{\rm RRL} = (0.08 -1) \times 10^{-3}$ through line stacking, reaching the range of optical depths observed from ionized gas in the Milky Way \citep[e.g.,][]{Roshi2000}. For an electron temperature of 8000~K, these optical depths convert into emission measures $10^{2.6} \gtrsim EM / \mathrm{cm^{-6}~pc} \gtrsim 10^{4.7}$ for densities $1 ~\mathrm{cm^{-3}} \lesssim n_e \lesssim 10^4$~\cmc{} (see Fig.~\ref{fig:qn_ne}).

The first science verification observations of MALS \citep{Gupta2021a, Combes2021} focused on the bright ($S_{1\,\mathrm{GHz}} \approx 11$~Jy) $z = 2.507$ \citep{Lidman1999} blazar, PKS\,1830$-$211 (referred to hereafter as \trg{}). \trg{} has a radio spectral index of $\alpha_{1-15~\rm GHz} \approx -0.26$, classifying it as a flat spectrum radio quasar \citep{PrameshRao1988, Subrahmanyan1990}, and at lower frequencies it is known to have a spectral turnover \citep{Lovell1996}. \trg{} is strongly gravitationally lensed \citep{Patnaik1993, Nair1993} by a galaxy at $z=0.89$ \citep{Wiklind1996}. Its morphology reveals two compact radio components, referred to as northeast (NE) and southwest (SW), approximately 1\asec{} apart and surrounded by a low surface-brightness Einstein ring \citep[e.g.,][]{Jauncey1991}. While the NE and SW components are images of the blazar core, the ring is mainly due to the jet and a bright knot in the jet \citep{Jin2003}. \trg{} is known to be variable, by up to a factor of two in radio, on timescales of hours to years \citep[e.g.,][]{PrameshRao1988, Marti-Vidal2013, Allison2017, Marti-Vidal2019}, and these variations are seen in all three lensed components --- the NE, SW, and ring.  The continuum flux is dominated by the NE and SW components at least down to 1.4~GHz \citep{Verheijen2001, Koopmans2005}, where the Einstein ring contributes $\sim$1\% to the continuum flux \citep{Combes2021, Patnaik1993}. For an overview image of the system, we refer the reader to \cite{Nair1993}.

Two absorption line systems are present along the line-of-sight to \trg{}. The lensing galaxy at $z=0.89$ has become an extragalactic-prototype absorption system, with the most molecular species detected to date (towards the SW image) \citep[e.g.,][]{Wiklind1996, Muller2011, Tercero2020}. The $z=0.89$ galaxy has been directly imaged with \textit{Hubble} Space Telescope and appears to be a barred-spiral \citep{Courbin1998, Courbin2002}, but remains weak and elusive. From a well-constrained lensing model \citep{Nair1993, Koopmans2005, Muller2020b, Combes2021}, the kinematics ($v_{rot} \sim 260$~\kms{}) and orientation also suggest a nearly face-on ($i \sim 25$\deg{}) barred-spiral galaxy of $\sim$10$^{11}$~\Msun{}. The SW image of the blazar core passes through the galaxy at a radius of $R_g \sim 2.4$~kpc and a central velocity (from spatially unresolved emission) at $v_{\rm cen} \sim 0$~\kms{}, and the NE image passes through at $R_g \sim 5.3$~kpc and $v_{\rm cen} \sim - 150$~\kms{}. 
The SW image intercepts a spiral arm of the $z=0.89$ galaxy and traces dense ($n_{H_2}\sim 2000$~cm$^{-3}$) molecular gas \citep{Wiklind1996, Courbin2002, Muller2013}. The gas along the line of sight to the NE image arises in the diffuse ISM \citep{Muller2011} and is bright in \Hi{} 21~cm and main OH 18~cm absorption \citep{Chengalur1999, Koopmans2005, Gupta2021a, Combes2021}. No time variation is visible in the cm-line spectra from the $z=0.89$ galaxy \citep{Combes2021}, which contrasts strongly with the variations detected in the mm-wave absorption spectra \citep{Muller2008, Muller2014a, Schulz2015}. The second absorption system towards \trg{} at $z=0.19$ has been seen only in \Hi{} 21 cm absorption so far \citep{Lovell1996, Allison2017, Gupta2021a}.

MALS observations of \trg{} verified system performance and led to the first detection of OH 18~cm satellite lines at $z = 0.89$, which had previously only been detected at $z \lesssim 0.25$ \citep{Gupta2021a, Combes2021}. In this article, we use these MALS observations to search for radio recombination line emission. We aim to understand whether RRLs are present and detectable and what they can tell us about photoionized gas in galaxies and AGN. In Sec.~\ref{sec:data}, we describe the methods used to process the data. In Sec.~\ref{sec:results}, we report (i) detections of hydrogen RRLs in both the L and UHF bands originating from ionized gas in the $z=0.89$ galaxy, (ii) tests we performed to verify these results, and (iii) non-detections at the additional redshifts searched. In Sec.~\ref{sec:physcond}, we constrain physical conditions of the ionized gas by modeling the stimulated RRL emission. Finally, we discuss implications of the ionized gas measured by the RRLs in Sec.~\ref{sec:discuss} and conclude in Sec.~\ref{sec:conclude}. In this article, velocities are reported using the heliocentric frame, with respect to $z=0.88582$ unless stated otherwise, and are converted from frequency using the relativistic definition. We use $\Lambda$CDM cosmology with $\Omega_m = 0.29$, $\Omega_{\Lambda} = 0.71$, and $H_o = 70$~\kms{}~Mpc$^{-1}$, for which 1\asec{}~$\sim$~7.860~kpc at $z=0.88582$.

%%%%%%%%%%%%%%%%%%%%%%%%%%%%%%%%%%%%%%%%

\section{Data and Processing}    
\label{sec:data}

%%%%%%%%%%%%%%%

\subsection{Observations and Data Reduction}

\trg{} was observed with the MeerKAT Radio Telescope \citep{Jonas2016} as the first science verification target of MALS \citep{Gupta2016}. 
For the work presented here, we used MALS L band spectra originally presented by \cite{Gupta2021a}. Hereafter, we refer to this L band dataset obtained on 2019 December 19 as ``Night1''. We also used UHF band spectra from the dataset obtained on 2020 July 13 and presented in \cite{Combes2021}. In addition to these previously published datasets, we observed \trg{} a second time in L band on 2020 September 18 using 59 antennas, which we will refer to as ``Night2'' in the article. 

For both L band observations, the total bandwidth of 856 MHz was centered at 1283.987~MHz, covering 856--1712~MHz, and split into 32\,768 frequency channels. This delivered a frequency resolution of 26.123~kHz, which is 6.1~\kms{} at the center of the band. For UHF band, the total observable bandwidth of 544~MHz covering 544--1088~MHz was also split into 32\,768 frequency channels, providing a channel resolution of 16.602~kHz, or 6.1~\kms{} at the center of the band, i.e., 815.992~MHz. For all observations the correlator dump time was 8~s and the data were acquired for all four polarization products, labeled as XX, XY, YX and YY. We also observed PKS\,1934--638 and/or 3C\,286 for flux density scale and bandpass calibrations. Since \trg{} is a bonafide VLA gain calibrator at this spatial resolution, a separate gain calibrator was not observed. The total on-source times on \trg{} are: 40\,min (L band Night1), 90\,min (UHF band) and 90\,min (L band Night2).

All MALS datasets have been processed using \artip{}, the Automated Radio Telescope Imaging Pipeline \citep{Gupta2021a}, a Python-based pipeline using tasks and tools from the Common Astronomy Software Applications \citep[CASA;][]{McMullin2007, CASATeam2022}. The specific details of the observations, calibration, and imaging of L and UHF band datasets can be found in \citet{Gupta2021a} and \citet{Combes2021}, respectively. The Stokes-$I$ continuum flux densities of \trg{} obtained from wideband radio continuum images in L Night1 and UHF band with {\tt robust=0} weighting are $11.245 \pm 0.001$~Jy at 1270~MHz and $11.40 \pm 0.01$~Jy at 832~MHz, respectively. The radio emission is unresolved in these images with a spatial resolution of $12\farcs9\times8\farcs1$ (position angle = $-76\fdg3$) and $17\farcs4\times13\farcs1$ (position angle = $+69\fdg0$), respectively \citep[][]{Gupta2021a, Combes2021}. The quoted uncertainty of the flux densities are based on a single 2D Gaussian fit to the continuum emission. The continuum flux density of \trg{} obtained using the Night2 dataset is $S_{1.27~\mathrm{GHz}} = 11.86 \pm 0.02$~Jy. This matches with the Night1 measurement within the flux density uncertainty ($\sim$5\%) expected at these low frequencies. Therefore, throughout the article we use the average flux density from the Night1 and Night2 datasets as the representative L band flux density, i.e., $S_{1.27~\mathrm{GHz}} \approx 11.5$~Jy.

The spectral line data products from \artip{} for RRL analysis are continuum-subtracted XX and YY parallel hand image cubes obtained with {\tt robust=0} weighting. We also note that for spectral line processing \artip{} splits the L and UHF bands into 15 spectral windows (hereafter SPWs) \citep[see][for details of SPW boundaries]{Gupta2021a, Combes2021}. The pixel sizes for L and UHF band image cubes are $2\farcs0$ and $3\farcs0$, respectively. XX and YY spectra were extracted in all SPWs from a single pixel at the location of \trg{} determined from the continuum images. The residual flux density in each continuum-subtracted spectrum is on the order of 0.5\% and required additional spectral-baseline subtraction. 

Further details of RRL specific spectral line processing are provided in the next section. In passing, we note that we also made use of a MALS L band dataset of another radio source, PKS~1740-517 (hereafter PKS\,1740), also known as J1744-5144, observed on 2020 September 20 (two days after Night2 observations of \trg{}) with an on-source time of 56\,min. This observation also used PKS\,1934-638 and 3C\,286 for flux density and bandpass calibrations, and the unresolved radio source has a flux density of $\sim$7\,Jy at 1270\,MHz, comparable to \trg{}. We process this dataset following the procedures described above and use the resultant spectra to establish the genuineness of the results obtained for \trg{}.

%%%%%%%%%%%%%%%

\subsection{RRL Spectral Processing}

\begin{figure*}
    \centering
    \includegraphics[width=\textwidth]{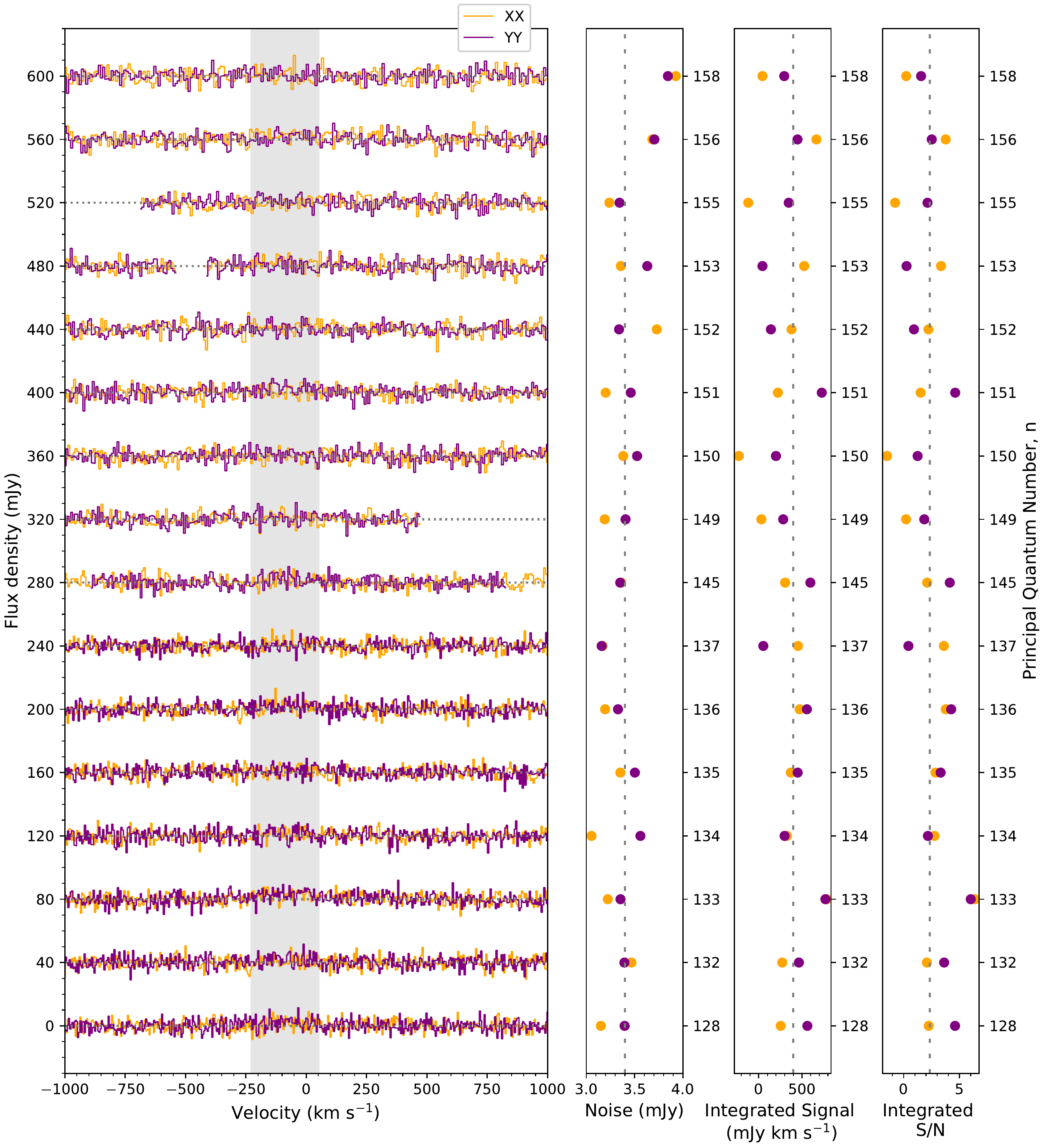}
    \caption{Spectral properties prior to stacking RRL transitions for L band spectra from Night 2. Velocities are shown with respect to $z=0.88582$ and spectra are shifted along the ordinate for display purposes. The principal quantum number of each spectrum is given on the right hand side ordinate. The shaded gray region in the left most panel indicates the line blank region. ``Noise'', $\sigma$, is the rms (outside of the line blank region) per 8~\kms{} channel of the XX (yellow circles) or YY (purple circles) spectrum. ``Integrated Signal'' $= \Delta v \Sigma_{i=0}^{N}\, S_i$, the sum of emission from the channels inside the line blank region. ``Integrated S/N'' $= (\Sigma_{i=0}^{N}\, S_i) / (\sqrt{N}\sigma)$, the integrated signal divided by the integrated noise; we note that corresponding to negative values of Integral Signal, the Integrated S/N is also negative. The vertical dashed line in the three right panels indicates the median value. There are no obvious and significant spurious features that could contaminate the RRL spectrum in our final stacking. }
    \label{fig:proc}
\end{figure*}

Considering the rest frequencies of RRLs, 38 and 44 of the $\alpha$ ($\Delta \mathsf{n} =1$) recombination lines (per element) fall within the MALS L and UHF band coverage, respectively.\footnote{We refer to the reader to the CRRLpy module \citep{Salas2016} found at \url{https://github.com/astrofle/CRRLpy} for a list of RRL line frequencies.} At $z=0.89$, for example, the observations cover 31 $\alpha$ recombination lines in L band spanning principal quantum numbers $\mathsf{n} = 128 - 158$ and 36 $\alpha$ recombination lines in UHF band spanning $\mathsf{n} = 148 - 183$. Because line properties of RRLs are correlated over a sizeable frequency range, we stacked the spectral lines to drive down the noise and increase the signal to noise ratio, as is commonly done in Galactic and extragalactic RRL observations \citep{Balser2006, Emig2020a}.

We began spectral processing by identifying the observed frequency of an RRL and extracting a spectrum equivalent to $v_{\mathrm{sys}} \pm 1500$~\kms{} centered on the line. We selected this velocity chunk and discarded coverage outside of it in order to (i) have a sufficient number of channels to minimize errors in the estimation of the spectral-baseline continuum level \citep{Sault1994}, while (ii) ensuring that a low ($\leq5$) order polynomial -- with an order determined by minimizing the reduced $\chi^2$ -- could be fit over a well-behaved bandpass.  RRLs that fell within $\pm 1000$~\kms{} from the edge of a spectral window were excluded from subsequent processing. We flagged channels with persistent radio frequency interference (RFI) and/or at the HI 21~cm and OH 18~cm line features (at relevant redshifts) \citep[see][]{Gupta2021a, Combes2021}, and we discarded the full spectral line chunk when flagged channels fell within $|v| \sim 500$~\kms{} from the line center in order to ensure a reliable fit to the spectral-baseline. We also flagged spectral line chunks that were clearly contaminated by broad-band RFI and reliable estimates of the baseline could not be attained. 

After flagging, we next fit a low order ($\leq5$) polynomial to (line-free) channels in each spectral chunk and subtracted the fit. For the results presented in Sec.~\ref{sec:results}, we did not impose a line-blank region, except for the $z=0.89$ stacks. For the $z=0.89$ results, we first stacked the spectrum without a line-blank region. Based on the significant feature we obtained in that spectrum, we set the line blank as -230~\kms{} to +55~\kms{}.

Fig.~\ref{fig:proc} shows the baseline-subtracted spectral chunks, as an example, from L band Night2 observations processed for $z=0.89$ RRLs. The spectral noise has a median of 3.4~mJy across the 17 RRL spectral chunks used in the stack. The noise properties and number of lines used in the final stacks can be found in Table~\ref{tab:line}. The noise properties are consistent across the bands and between parallel hand spectra.

We next interpolated each spectral chunk to a common velocity grid with channel widths of 1~\kms{}, intentionally oversampling the spectral channels to avoid artificially smoothing-out spectral features. The spectral line chunks were then averaged together using the inverse noise squared in line-free channels as a weight \citep[e.g.,][]{Emig2020a}.\footnote{ We tested combining the spectral chunks with an additional weight that depended upon the (inverse) line frequency (raised to a power), but this did not significantly change the S/N properties, indicating that the line properties are similar and well-correlated across the observing bands.} We next smoothed the channel resolution to 8~\kms{} using a boxcar averaging kernel, to better match the lowest resolution achieved at the low frequency end of the bands. At this stage, we had obtained a single H$n\alpha$ spectrum for each parallel hand XX and YY. Finally, we averaged the XX and YY spectra to create a Stokes-$I$ spectrum. We chose to combine the stacked XX spectrum and the stacked YY spectrum rather than creating a Stokes-$I$ spectrum for each line before combining polarizations because it (i) resulted in better-behaved, i.e., Gaussian-like noise properties and (ii) had the benefit of independently examining the differences in the signal detected in two parallel hand spectra. The latter is particularly useful in distinguishing true astrophysical signal from RFI, which is often linearly polarized.

%%%%%%%%%%%%%%%%%%%%%%%%%%%%%%%%%%%%%%%%

\section{Results}   
\label{sec:results}

We applied the spectral processing procedures described in the previous section to \trg{} observations at the redshifts (i) $z=0$ for Galactic emission, (ii) $z=0.19259$ for the low redshift intervening absorber, (iii) $z=0.88582$ for the high redshift intervening absorber, and (iv) $z=2.507$ for the intrinsic redshift of \trg{}. We detect RRL emission from the $z=0.89$ absorber in both Nights of L band and in UHF band observations. We report non-detections and upper limits at all other redshifts and bands. 

% table: line properties 

\begin{deluxetable*}{lllccCRCCCC}
\tablecaption{Spectral and Line Properties \label{tab:line} }
\tablewidth{0pt}
\tablehead{
\colhead{Line-of-sight} %1
& \colhead{$z$} %2
& \colhead{Band} %3
& \colhead{$N_{\rm lines}$} %4
& \colhead{RRL} %5
& \colhead{noise} %6
& \colhead{$v_{\mathrm{center}}$} %7
& \colhead{$S_{\rm peak}$} %8
& \colhead{FWHM} %9
& \colhead{$\int S_{\rm RRL} \, \mathsf{d}v$} %10
& \colhead{$\int \tau \, \mathsf{d}v$} %11
\\
\colhead{} %1
& \colhead{} %2
& \colhead{} %3
& \colhead{} %4
& \colhead{} %5
& \colhead{(mJy)} %6
& \colhead{(\kms)} %7
& \colhead{(mJy)} %8
& \colhead{(\kms)} %9
& \colhead{(mJy \kms)} %10
& \colhead{(\kms)} %11
}
\startdata
PKS 1830-211    & 0.88582   & L & 17    & H\,144\,$\alpha$    & 0.34  & -117.4 \pm 5.3    & 1.86 \pm 0.12 & 131 \pm 14    & 258 \pm 33    & -0.045 \pm 0.006\\
                &   &   &   &   &   & 7.7 \pm 4.2   & 1.54 \pm 0.19     & 62.3 \pm 9.8  & 102 \pm 20  & -0.018 \pm 0.003 \\
                &   & UHF   & 27    & H\,163\,$\alpha$    & 0.57  & -124.4 \pm 9.4       & 1.59 \pm 0.15    & 155 \pm 24 & 262 \pm 47  & -0.046 \pm 0.008 \\
                &   &   &   &   &   & 7.7  & 0.81 \pm 0.25     & 80 \pm 28     & 70 \pm 25   & -0.012 \pm 0.004 \\
    & 0.0       & L & 17    & H\,175\,$\alpha$    & 0.36    & ...    & ...    & ...    & <22.7   & <|0.0020| \\
    &           & UHF & 28    & H\,203\,$\alpha$    & 0.41    & ...   & ...   & ...   & <26.5   & <|0.0023| \\
    & 0.19259   & L & 17    & H\,166\,$\alpha$    & 0.34   & ...   & ...   & ...    & 23.6 \pm 7.3 & - 0.002\,1 \pm 0.000\,6 \\
    &           & UHF   & 33    & H\,189\,$\alpha$    & 0.43   & ...   & ...   & ...   & <27.2   & <|0.0024| \\    
    & 2.507     & L & 14    & H\,116\,$\alpha$    & 0.40  & ...   & ...   & ...   & <25.5   & <|0.0023| \\
    &           & UHF   & 19    & H\,132\,$\alpha$ & 0.55  & ...   & ...   & ...   & <34.9   & <|0.0031| \\
PKS 1740-517   & 0.88582   & L  & 17    & H\,142\,$\alpha$ & 0.39  & ...   & ...   & ...   & <24.8 & <|0.004| \\
\enddata
\tablecomments{ Uncertainties of the line properties are determined from the variance of each parameter as determined by the fit. $N_{\rm{lines}}$ is the effective number of recombination lines stacked in the final spectrum. ``RRL'' refers to the effective radio recombination line transition of the stacked spectrum. ``noise" is the weighted standard deviation of line-free channels in units of mJy per 8~\kms channel. $\rm{v_{center}}$ is the central velocity of the best fit Gaussian and uncertainty. $S_{\rm{peak}}$ is the peak amplitude of the best fit Gaussian fit. FWHM is the full-width half maximum of the best fit Gaussian. $\int S_{\rm RRL} \, \mathsf{d}v$ is the velocity-integrated flux density of the best-fit Gaussian profile, or in the case of an upper limit, the reported value is equal to an integrated signal to noise ratio of 3 assuming a line width of 60~\kms. $\int \tau \, \mathsf{d}v$ is the velocity-integrated optical depth computed as $- \int S_{\rm RRL}/ S_{\rm c} ~ \mathsf{d}v$ where the values used for the continuum flux density, $S_{\rm c}$, are described in Sec.~\ref{sec:results}. }
\end{deluxetable*}

\begin{figure}
    \centering
    \includegraphics[width=0.47\textwidth]{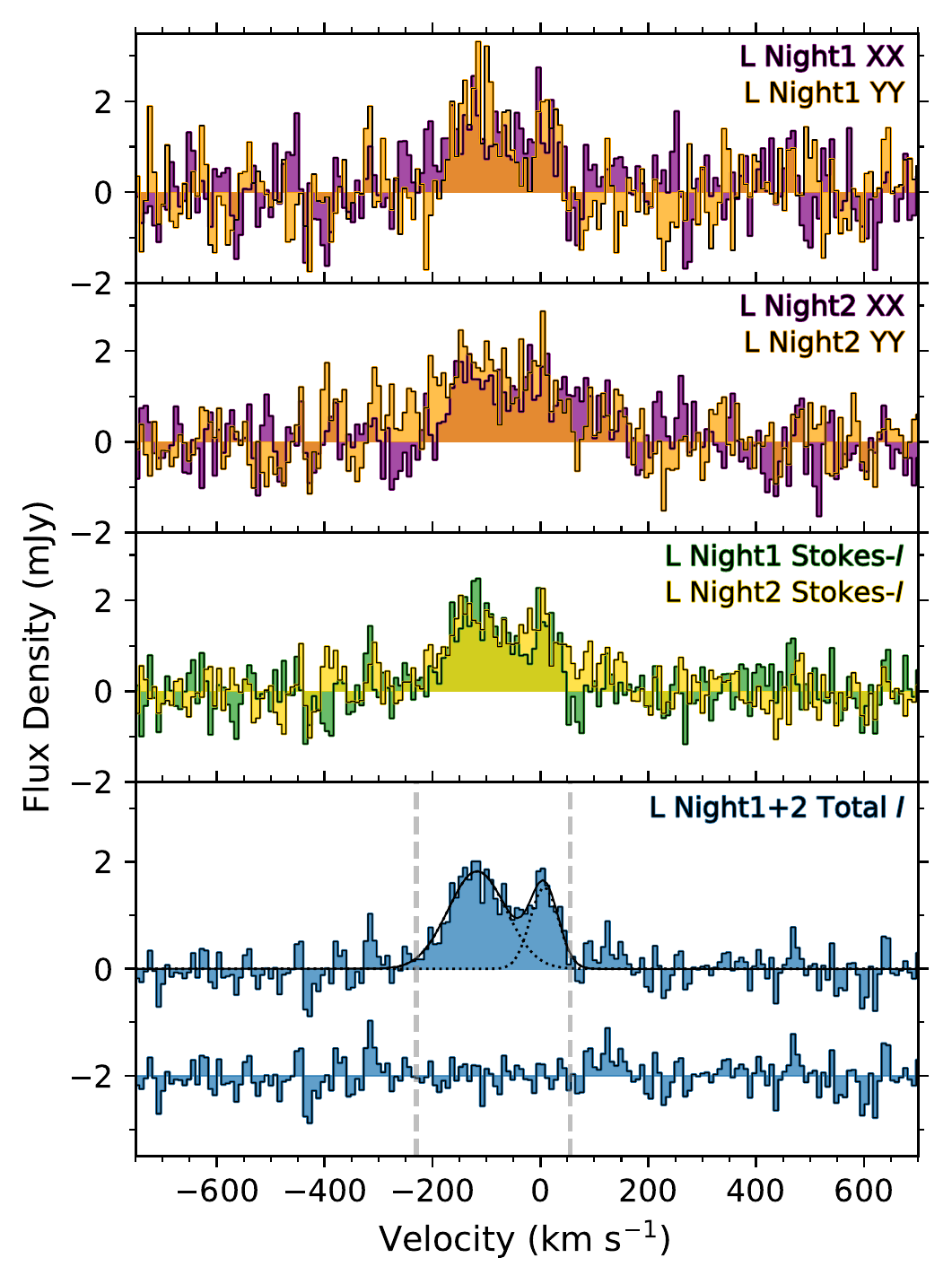}
    \caption{Comparison of RRL stacked spectra in L band at $z=0.89$. The top three panels compare spectra from parallel hands and observing nights. The bottom panel shows (i) the final spectral result for the band with Gaussian profiles fit to two components and (ii) underneath, the final spectrum with the Gaussian profiles subtracted. The vertical dashed lines mark the velocities $-230$~\kms{} and 55~\kms{} within which we integrate the signal of the Total component.}
    \label{fig:spec_z0p89_L}
\end{figure}

In Fig.~\ref{fig:spec_z0p89_L}, we show the L band detections obtained from the $z=0.89$ absorber. We overlay the parallel hand spectra, showing that they are consistent within the noise in both nights of observations and therefore not due to low-level linearly-polarized RFI. We also overlay the Stokes-$I$ spectrum obtained from each night of L band observations; they are consistent within the noise, giving further evidence that the signal is astrophysical in nature.  
Lastly, we averaged L band Stokes-$I$ spectra from Nights 1 and 2, producing a spectrum with the highest signal-to-noise ratio attainable which we refer to as the total-$I$ spectrum. The reduction in spectral noise of the incrementally combined spectra follows root N statistics. The total integrated flux density of $\int S_{\rm H144\alpha} \, \mathsf{d}v = 337 \pm 16$~mJy~\kms{} with a maximum integrated signal-to-noise ratio (S/N) $= (\Sigma_{i=0}^{N}\, S_i) / (\sqrt{N}\sigma)$ of 21 is computed from the total-$I$ spectrum by integrating over the $N$ channels covering the velocity range $-230$ to 55~\kms{}. The effective transition of the total-$I$ spectrum presented in the bottom panel of Fig.~\ref{fig:spec_z0p89_L} is H144$\alpha$, at a rest-frame frequency of 2179.6~MHz and observed at 1155.8~MHz. The effective transition is determined from the noise-weighted average of the line transitions included in the stack. The line properties of the observed emission are consistent with $S_{\mathrm{H}136\alpha} < 5$~mJy upper limits to the H136$\alpha$ line at $\nu_{\mathrm{rest}} = 2585.7$~MHz ($\nu_{\mathrm{obs}} = 1371.1$~MHz) --- a transition which is included in our stack --- obtained by \cite{Mohan2002a}.

% figure: spectra of detections

\begin{figure}
    \centering
    \includegraphics[width=0.47\textwidth]{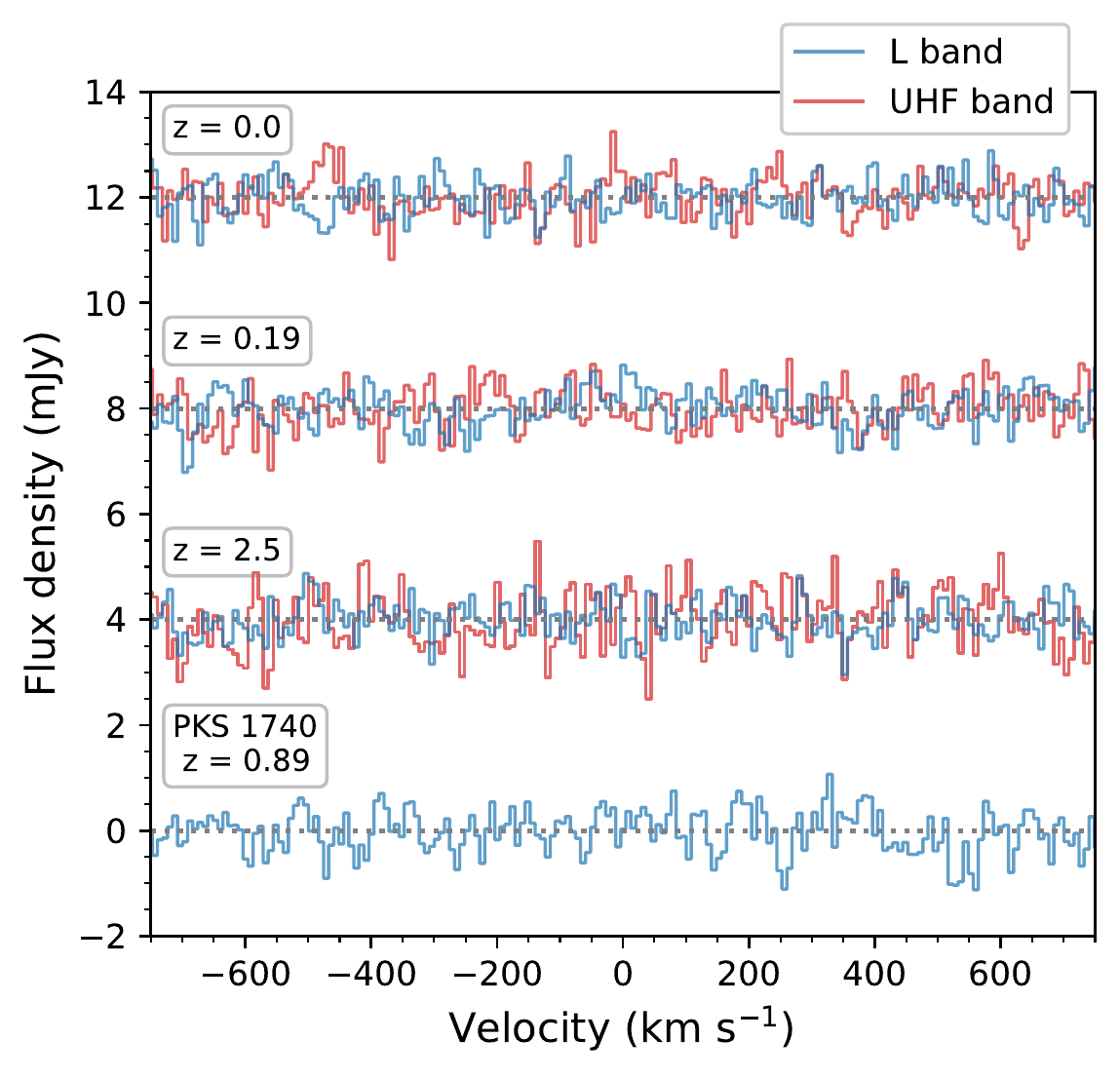}
    \caption{Non-detection RRL spectra in the spectrum of \trg{} (top three) and in PKS\,1740 (bottom). Spectra have been given an arbitrary offset in intensity in multiples of 4~mJy. Velocities are defined with respect to the labeled redshift.}
    \label{fig:spec_nondetect}
\end{figure}

In addition to (1) multiple nights of observations and (2) comparing XX and YY parallel hand spectra, we verified additional evidence that the signal is true recombination line emission by (3) observing an independent detection of RRL emission in UHF band (more details below), (4) performing jackknife tests, in which one line spectrum at a time is omitted from the stacked spectrum, showing that the signal is not dominated by a single outlying spectral chunk\footnote{We also refer the reader to Figure~\ref{fig:proc}, where the noise, integrated signal, and integrated signal-to-noise properties also demonstrate no single or few outlying spectra dominate.}, (5) splitting the lines in the band into two groups creating two sub-stacks and this resulted in consistent line properties, (6) verifying a signal is not reproducible by stacking RRLs at other redshifts (more details at the end of the Section and see Fig.~\ref{fig:spec_nondetect}), and (7) finding that an RRL spectrum of PKS\,1740 stacked at $z=0.89$ does not systematically produce a signal. We made use of additional MALS L band observations of PKS\,1740 and followed the spectral processing procedures described in Sec.~\ref{sec:data}. We created an average RRL spectrum at $z=0.89$ with an effective transition of H142$\alpha$ and show it in Fig.~\ref{fig:spec_nondetect}. This spectrum reached a noise, $\sigma = 0.39$~mJy, comparable to the \trg{} stack. However, no emission or significant spectral features are present in the PKS\,1740 spectrum, and it is consistent with noise. This lends additional support to the physical and real nature of the emission from the $z=0.89$ absorber in \trg{}.

% figure: spectra of non-detections

Two velocity components dominate the L band H144$\alpha$ emission from \trg{}. In the bottom panel of Fig.~\ref{fig:spec_z0p89_L}, we show the best fit of two Gaussian profiles and the spectrum that results when the two Gaussian profiles are subtracted. The properties of the Gaussian fits are listed in Table~\ref{tab:line}; the errors of each quantity are determined from the variance of each variable as determined by the fit. The H144$\alpha$ component centered on $-117.4 \pm 5.3$~\kms{} arises from the NE sight-line (and thus we will refer to this velocity component as the NE component hereafter). The H144$\alpha$ component centered on $7.7 \pm 4.2$~\kms{} arises from the SW sight-line (and thus we will refer to this velocity component as the SW component hereafter). In Table~\ref{tab:line}, we also include the integrated optical depth equal to $\int \tau \, \mathsf{d}v \approx - \int S_{\mathrm{RRL}} / S_{\rm c} ~ \mathsf{d}v$ computed by letting $S_c$ of each component equal half the total continuum flux density in the band, $S_c \approx 5.75$~Jy. We assume the continuum flux is equally split between the two core components following \cite{Koopmans2005} and \cite{Combes2021} which show this to be the case and that the core components dominate the emission at least down to 1.4~GHz. Furthermore, ALMA observations measure a NE/SW flux density ratio close to one in July 2019 \citep{Muller2021}. 

\begin{figure}
    \centering
    \includegraphics[width=0.47\textwidth]{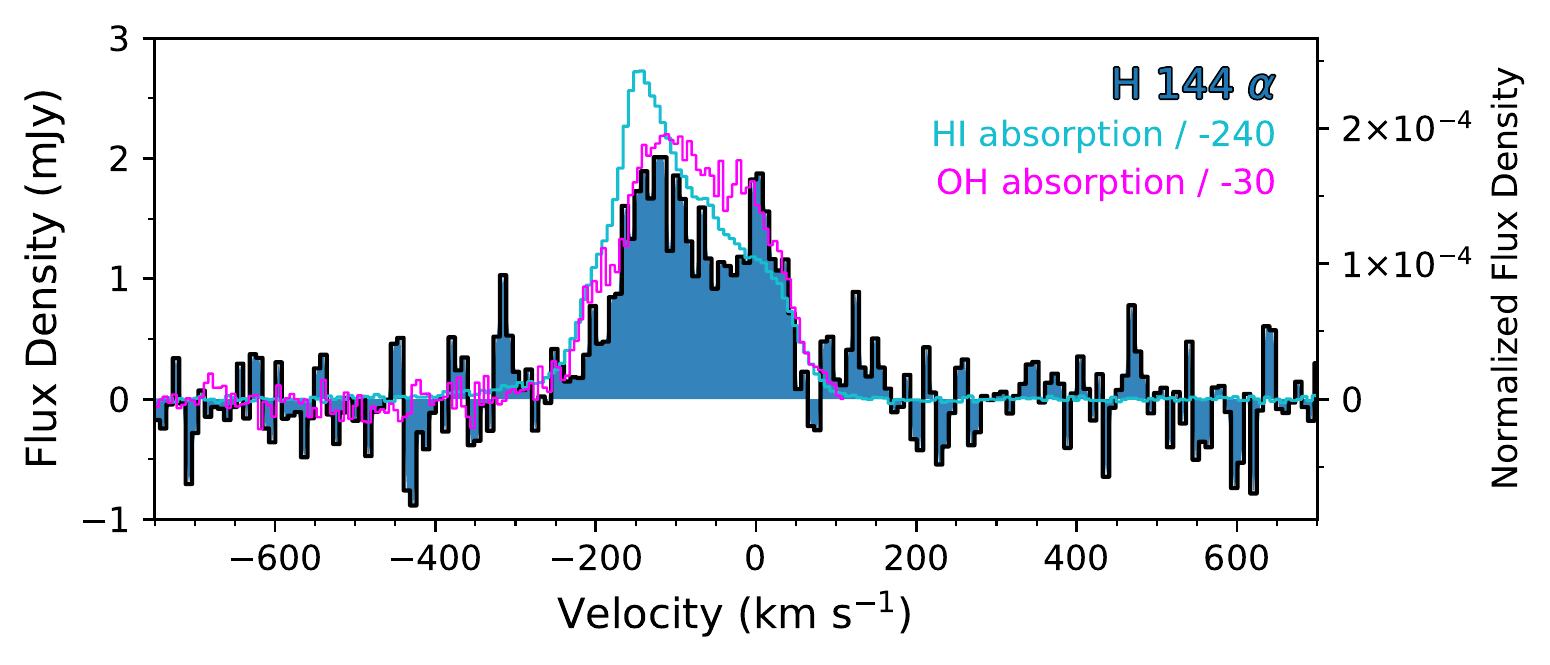}
    \caption{The H144$\alpha$ spectrum detected in L band overlaid by the rescaled \Hi{} and OH absorption spectra. The right-hand ordinate shows the RRL flux density normalized by the continuum.}
    \label{fig:compare_tracers}
\end{figure}

Fig.~\ref{fig:compare_tracers} overlays the H144$\alpha$ spectrum with the \Hi{} 21~cm and OH 18~cm absorption spectra that have been rescaled by factors of $-240$ and $-30$, respectively, in flux density units and obtained from the same MALS datasets. The RRL emission appears to span a similar velocity range as these other diffuse gas tracers and likewise, is also dominated by two main velocity components. 
    
The RRL line centroids of the NE and SW components agree within error with the OH 18~cm profiles, which have values of $-110 \pm 3$~\kms{} and $6 \pm 3$~\kms{}, respectively. However, the RRL centroids are significantly offset from the dominant \Hi{} components at $\sim -150$~\kms{} and 0~\kms{}, respectively, albeit the \Hi{} profile is more complex with perhaps five velocity components best-fitting the observed profile. The variation in central velocity with \Hi{} could arise due to different filling factors, intrinsic distributions, and shapes of the continuum with the higher frequency tracers (OH and RRLs) finding better agreement. 

The RRL line width of the NE component also agrees within error with the OH absorption width, but the widths are significantly different for the SW component, with the RRL FWHM of $63 \pm 10$~\kms{} and the OH FWHM of $94.2 \pm 5$~\kms{}. An estimate of the \Hi{} width is close to $\sim 100$~\kms{} for both components, which would be inconsistent with the RRL widths from either component. Given the smaller line width of the warmer gas traced by the RRLs in the SW, the emission may likely have a smaller filling factor, which is to say, fewer individual components contribute to the total profile. This can be expected since the SW line of sight is dominated by dense molecular gas.

While the two dominant components of H144$\alpha$ emission generally agree with the \Hi{} and OH profiles, more than two velocity components are discernible in the higher S/N spectra of \Hi{} and OH. For example, OH absorption has an additional component centered at $-211 \pm 3$~\kms{} with a FWHM of $28 \pm 9$~\kms{}; the RRL line profile fit to the NE component encompasses some emission in this velocity range. The \Hi{} absorption spectrum also shows two velocity features that are blueshifted with respect to the main $\sim -150$~\kms{} peak.

\begin{figure}
    \centering
    \includegraphics[width=0.47\textwidth]{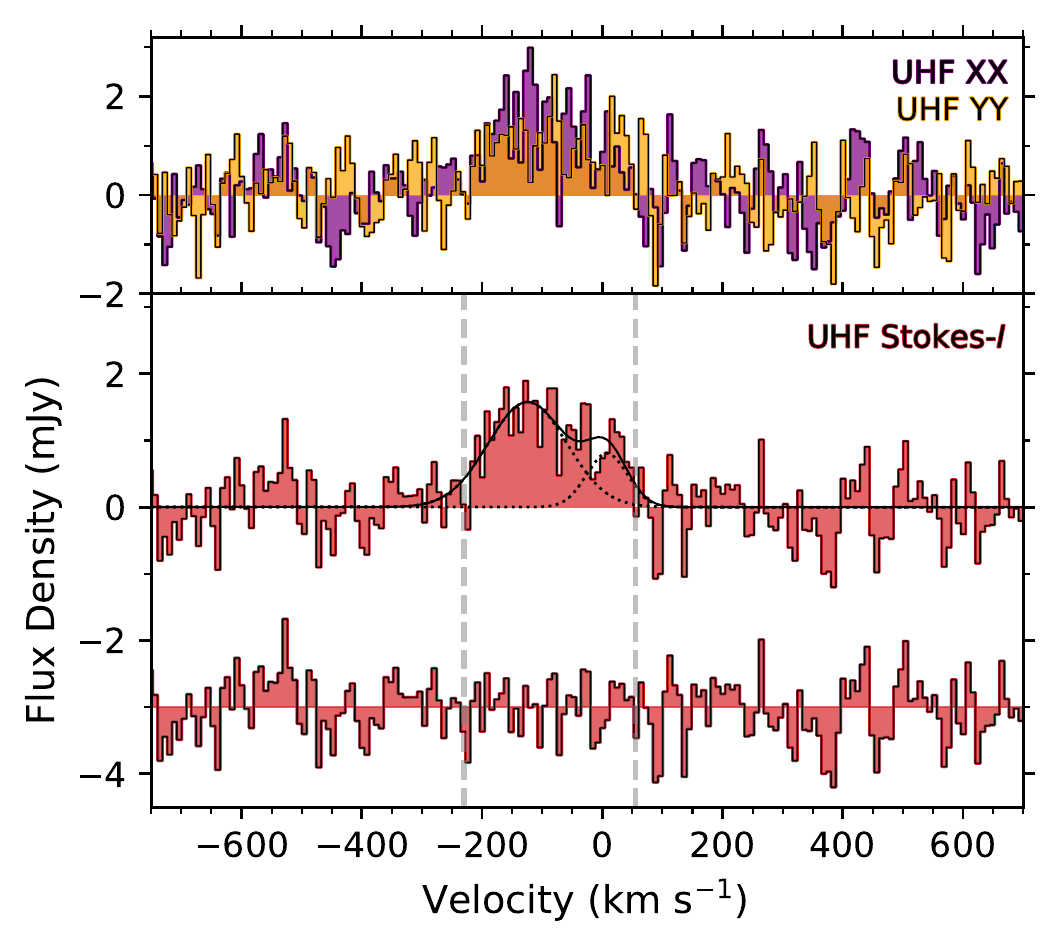}
    \caption{RRL stacked spectra in UHF band at $z=0.89$. The top panel compares spectra from the two parallel hands. The bottom panel is the same as in Fig.~\ref{fig:spec_z0p89_L} but for the UHF band spectrum.}
    \label{fig:spec_z0p89_UHF}
\end{figure}

In Fig.~\ref{fig:spec_z0p89_UHF}, we show the UHF band detections obtained at the $z=0.89$ absorber. We overlay the parallel hand spectra, showing that they are consistent within the noise and thus likely not a result of RFI. Integrating the Stokes-$I$ spectrum over the velocity channels from $-230$ to 55~\kms{} results in $\int S_{\rm H163\alpha} \, \mathsf{d}v = 309 \pm 22$~mJy~\kms{} and a maximum integrated S/N of 14. The effective transition of the final spectrum is H163$\alpha$, with a rest-frame frequency of 1504.6~MHz and observed at 797.85~MHz. 

The H163$\alpha$ emission arises across a similar velocity range as the H144$\alpha$ emission, yet the peak intensities are slightly lower. The distinction of two components is less obvious in the H163$\alpha$ stack (UHF band), as compared with the H144$\alpha$ stack (L band). We fit two Gaussian components to the spectrum, fixing the central velocity of the SW component equal to that obtained from the high S/N spectrum in L band, $v_{\mathrm{cen}} = 7.7$~\kms{}. The best fits are shown in Fig.~\ref{fig:spec_z0p89_UHF} and the fit parameters are listed in Table~\ref{tab:line}. As in the L band spectrum, we compute the integrated optical depth by assuming the UHF band continuum flux is equally split towards each core component, $S_c \approx 5.7$~Jy.

% figure: spectra of detections

In Fig.~\ref{fig:spec_nondetect}, we show the L and UHF band stacked spectra of \trg{} at each redshift where we obtained null results: $z=0$, $z=0.19$, and $z=2.5$. The spectral properties of the non-detections are included in Table~\ref{tab:line} and the integrated optical depth, $\int \tau \, \mathsf{d}v \approx - \int S_{\mathrm{RRL}} / S_{\rm c} ~ \mathsf{d}v$, is  computed by letting $S_c$ equal the total continuum flux density in the L and UHF bands, respectively. \cite{Mohan2002a} previously reported a 5$\sigma$ upper limit to the H158$\alpha$ line from the $z=0.19$ absorber of $S_{\rm H158\alpha} < 0.5$~mJy. In our L band spectrum of the $z=0.19$ stack at an effective transition of H166$\alpha$, we reach a spectral noise of 0.34 mJy, and our 3$\sigma$ upper limit to the peak line emission of $S_{\rm H166\alpha} < 1.0$~mJy is consistent with the results obtained by \cite{Mohan2002a}. 

Lastly, we note that carbon RRL emission becomes significantly enhanced only at frequencies $\lesssim$ 350 MHz, and thus we do not expect to detect it at the frequencies of our observations. The detected signal likely arises only from hydrogen emission given that it is coincident in velocity with the HI 21~cm and OH 18~cm absorption. A 3$\sigma$ upper limit to the carbon RRLs at $z=0.89$ in L band is $\int S_{\rm C144\alpha}\, \mathsf{d}v < 22.3$~mJy~\kms{} and in UHF band is $\int S_{\rm C163\alpha}\, \mathsf{d}v < 37.5$~mJy~\kms{}, assuming a line width of 60~\kms{}. We searched for H$\beta$ emission by stacking all available lines in both L and UHF band following the procedures described in Sec.~\ref{sec:data}. We report a 3$\sigma$ upper limit to the H$\beta$ emission of $\int S_{\rm H192\beta}\, \mathsf{d}v < 18.3$~mJy~\kms{} and an in-band ratio of peak H192$\beta$/H144$\alpha <0.5$, where typical in-band $\beta/\alpha$ ratios are $\sim 0.2$ \citep[e.g.,][]{Salas2017}.

%%%%%%%%%%%%%%%%%%%%%%%%%%%%%%%%%%%%%%%%

\section{Physical Conditions of Ionized Gas in the z=0.89 absorber}
\label{sec:physcond}

Because \trg{} has a relatively low Galactic latitude and is behind the Inner Galaxy, $(\ell, b = 12.0^{\circ}, -5.7^{\circ})$, it has not yet been feasible to observe ionized gas tracers at UV through IR wavelengths from the $z=0.89$ galaxy  \citep[e.g.,][]{Djorgovski1992, Courbin1998}. However, there have been some indications for the presence of ionized gas that could be attributed to the $z=0.89$ galaxy. 
Firstly, the jet emission which forms an Einstein ring shows a complete turnover at $\sim300$~MHz in its spectral energy distribution (SED), which could be due to free-free absorption and would not arise from synchrotron self absorption
\citep[][and for more details, see Sec~\ref{ssec:radioSED}]{PrameshRao1988, Jauncey1991, Jones1996, Lovell1996}.
Secondly, jet emission in the Einstein ring (i.e., not the core emission) is strongly polarized \citep{PrameshRao1988, Subrahmanyan1990}, indicating ionized plasma lies along the lines of sight.  However, it is debated whether the ionized gas originates in the Milky Way or the $z=0.89$ galaxy and if the blazar core components are free-free or synchrotron-self absorbed   \citep{Jones1996, Guirado1999, Marti-Vidal2015}.

Stimulated hydrogen radio recombination lines provide strong constraints on the gas volume density of electrons and volume-averaged pathlength. In the following sections we model the RRL emission to derive physical properties. We use these constraints to estimate the ionized gas mass per unit area and the ionizing photon flux.

%%%%%%%%%%%%%%%%%%%%%%%%%%%%%%%%%%%%%%%%

\subsection{RRL line width}
\label{ssec:rrlwidth}

The width of recombination lines as a function of frequency provides insight into the physical properties of the emitting gas. A Doppler-broadened profile has a constant width in velocity units as a function of frequency, and its Gaussian profile indicates broadening from the intrinsic gas particle motions (e.g., due to the temperature of the gas or turbulence) or from multiple emitting regions rotating at different velocities in a galaxy. 
However, collisional and radiation broadening create a Lorentzian line profile and have an increasing line width towards lower frequency, thereby informing on the electron density of the gas or the incident radiation field strength, respectively.

For the NE component, widths of FWHM$_{\rm H144\alpha} = 131 \pm 14$~\kms{} and FWHM$_{\rm H163\alpha} = 155 \pm 24$~\kms{} are consistent within error. For the SW component, widths of FWHM$_{\rm H144\alpha} = 62.3 \pm 9.8$~\kms{} and FWHM$_{\rm H163\alpha} = 80 \pm 28$~\kms{} are also consistent within error. This indicates that the lines are Doppler broadened. Gaussian distributions do fit our observed line profiles reasonably well (see Figs.~\ref{fig:spec_z0p89_L}~and~\ref{fig:spec_z0p89_UHF}).

Assuming the line widths of each component are equal at the two frequencies, the weighted average of the FWHM for the NE component is 137~\kms{} and for the SW component is 64~\kms{}. We then use the width at the lowest frequency to put a firm upper limit on the gas density, assuming pressure broadening dominates. Note, there is no indication to expect an extreme radiation field that would cause the line to be radiation broadened. \cite{Salgado2017b} provide the FWHM of a collisionally broadened profile, $\Delta \nu_{col} = n_e \mathsf{n}^{\gamma} \cdot 10^a / \pi $ where $\Delta \nu_{col}$ is in units of Hz, $n_e$ is in units of \cmc{}, $a = -7.386$, $\gamma = 4.439$ and $\mathsf{n}$ is the principal quantum level. We place an upper limit of $n_e \lesssim 7900$~\cmc{} for the NE component and $n_e \lesssim 3700$~\cmc{} for the SW component.

%%%%%%%%%%%%%

\subsection{RRL SLED}
\label{ssec:rrlmodel}

As shown in \cite{Shaver1975a} and \cite{Shaver1978b}, the flux density of an optically thin RRL of principal quantum number, $\mathsf{n}$, and frequency, $\nu$, observed in front of a significantly-brighter background radio source of flux density $S_{\rm{bkg},\nu}$ and which results from stimulated emission is given by
\begin{equation}
    S_{\mathsf{n},\nu} \approx - \tau^*_{\mathsf{n},\nu} \left( b_{\mathsf{n}}\beta \right) S_{\rm{bkg},\nu}
\label{eq:Sline}
\end{equation}
where $\tau^*_{\mathsf{n},\nu}$ is the LTE RRL optical depth, $b_{\mathsf{n}}$ is the ratio of the number of hydrogen atoms in level $\mathsf{n}$ between the non-LTE and the LTE cases, and $\beta$ characterizes the effect of stimulated transitions. $b_{\mathsf{n}}$ and $\beta$, collectively referred to as departure coefficients, depend on the atomic physics of the hydrogen atom, and their product is dependent upon electron density, electron temperature, the radiation field, and pathlength. Computation of the departure coefficients has been thoroughly studied since the first observations of hydrogen RRLs \citep[e.g.,][]{Shaver1975a, Hummer1987, Salgado2017a, Prozesky2018}. Integrating over the line profile and expressing the LTE optical depth of the line, the integrated optical depth of an $\alpha$ transition ($\Delta \mathsf{n} = 1$) at high principal quantum numbers takes the form
\begin{eqnarray}
\label{eq:inttau}
    \int \tau_{\mathsf{n}} \, \mathsf{d}v && \approx 6.13 \times 10^{-4} \,\mathrm{km\, s^{\text{-}1}} \left( b_{\mathsf{n}}\beta \right) \\ \nonumber 
     &&\left( \frac{EM}{10^4\, \mathrm{cm^{\text{-}6}\, pc}} \right) 
     \left( \frac{T_e}{10^4\, \mathrm{K}} \right)^{\text{-}2.5} 
    \left( \frac{\nu}{\rm{GHz}} \right)^{\text{-}1}
\end{eqnarray}
where $T_e$ is the electron temperature and $EM$ is the emission measure equal to $EM = \int n_e n_{ion} \, \mathsf{d}\ell$ for electron density $n_e$ and ion density $n_{ion}$ integrated over the pathlength $\ell$. Because the departure coefficients at each principal quantum number are strongly dependent upon density, they are key to breaking the degeneracy between density and pathlength in the emission measure.

\begin{figure}
    \centering
    \includegraphics[width=0.48\textwidth]{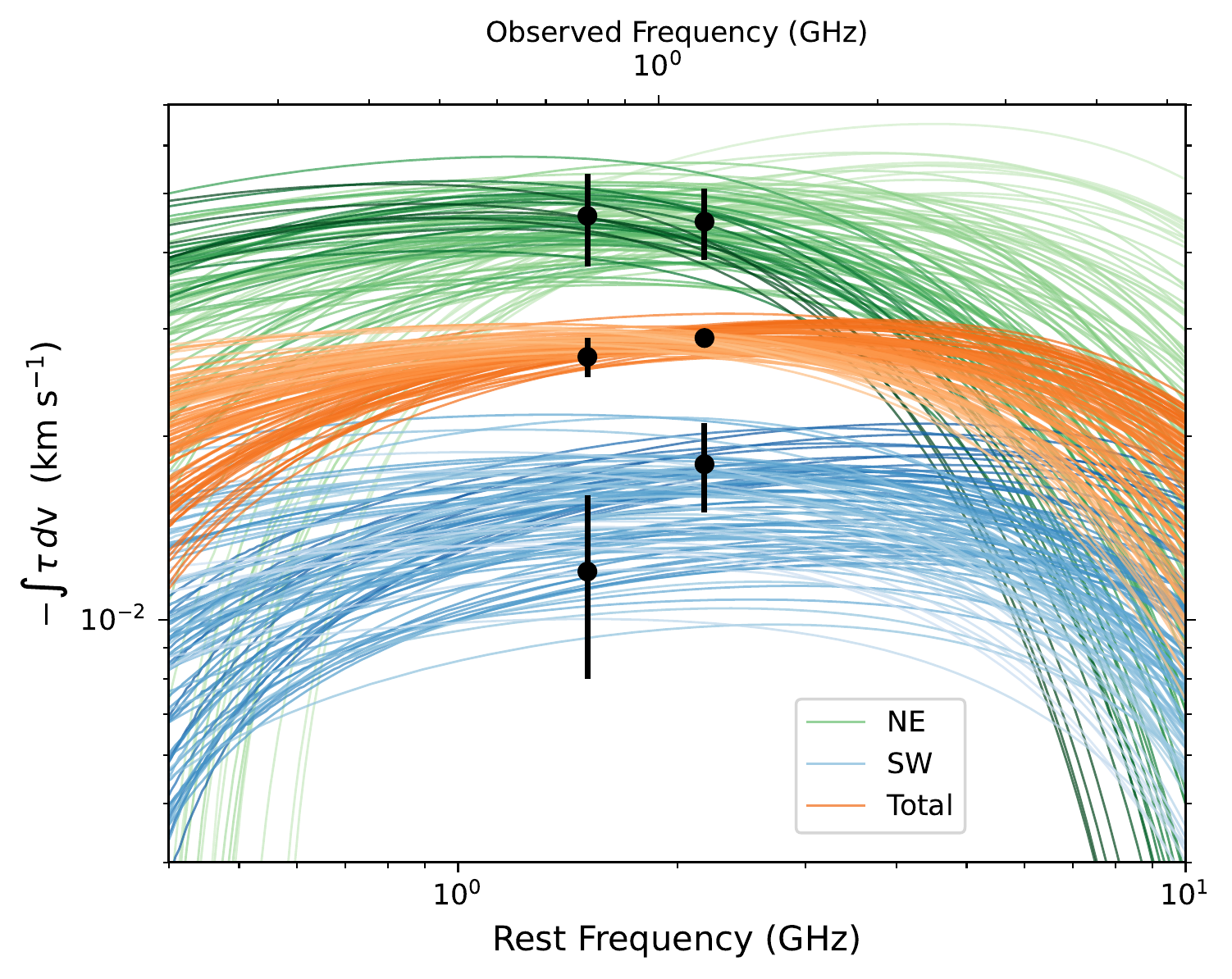}
    \caption{The integrated optical depth as a function of frequency for the NE, SW, and Total components. We overlay predicted line intensities from RRL models, from 100 model combinations that fall within the reported uncertainties and chosen at random.  We assume a redshift $z=0.89$ for the conversion between observed and rest frequencies. }
    \label{fig:SLED}
\end{figure}

In Fig.~\ref{fig:SLED}, we plot the RRL SLED using the integrated optical depths corresponding to the NE and SW Gaussian components from from Table~\ref{tab:line}. We also plot (and use for analysis in this section), the integrated optical depth of the Total component computed from the velocity-integrated emission over -230 to 55 \kms{} and setting the continuum flux density, $S_c$, equal to the flux density in each respective band, i.e. $\int \tau_{\rm{H}144\alpha} = - 0.029 \pm 0.001$~\kms{} and $\int \tau_{\rm{H}163\alpha} = - 0.027 \pm 0.002 $~\kms{}. While it would be ideal to model the two velocity components individually, the large errors of the Gaussian fit parameters warrant caution. The results from the Total component represent an average of the two lines-of-sight.

To model the recombination line emission, we calculated the departure coefficients for a range of electron densities and electron temperatures using the code and framework described in \cite{Salgado2017a, Salgado2017b}. When computing $b_{n} \beta $ we only consider the effect of the cosmic microwave background on the level populations. The grid in density was sampled at an interval of 0.5 dex and the temperature in multiples of 100. We then interpolate over the grid values using a cubic bivariate spline for the following analysis.

We used Bayesian analysis to constrain the posterior distribution of the parameters $n_e$, $T_e$, and $\ell$, and we assume the gas is fully ionized with $n_{ion} = n_e$. The likelihood function describes the comparison of the observed integrated optical depth with the model (Eq.~\ref{eq:inttau}), assuming a Gaussian distribution function with dispersion equal to the measurement uncertainty (see Table~\ref{tab:line}). The model is taken to be Eq.~\ref{eq:inttau} in which the values of $\nu$, $n_e$, $T_e$, and corresponding $b_{n} \beta$ are input, and the pathlength is left as a free parameter. We used flat priors for the parameters expressed in logarithmic scale. We assumed reasonable ranges of the parameters: for density, $\log(n_e) =[-1, 6]$ \cmc{} for the Total component, and we used the upper limits derived from the line width for the NE and SW components, $\log(n_e) =[-1, 3.4]$ and $\log(n_e) =[-1, 3.6]$, respectively, in units of \cmc{}; for pathlength, $\log(\ell) =[-9, 6]$ pc; and for temperature, we select a strict range of $\log( T_e ) =[3, 4.3]$ K, given that theory and observations substantiate temperatures of photoionized gas to have typical values of $T_e \approx 6000 - 10\,000$~K \citep[e.g.,][]{Wenger2019, Tielens2005}. We also used the prior that the departure coefficients must be negative and thus the line appears in emission. To sample the posterior distribution we used an affine-invariant sampler within the package \texttt{emcee} \citep{Foreman-Mackey2013}. We used 20 walkers, $10^5$ iterations, and verified convergence using an autocorrelation time analysis.

% figure of RRL physical properties
\begin{figure}
    \centering
    \includegraphics[width=0.48\textwidth]{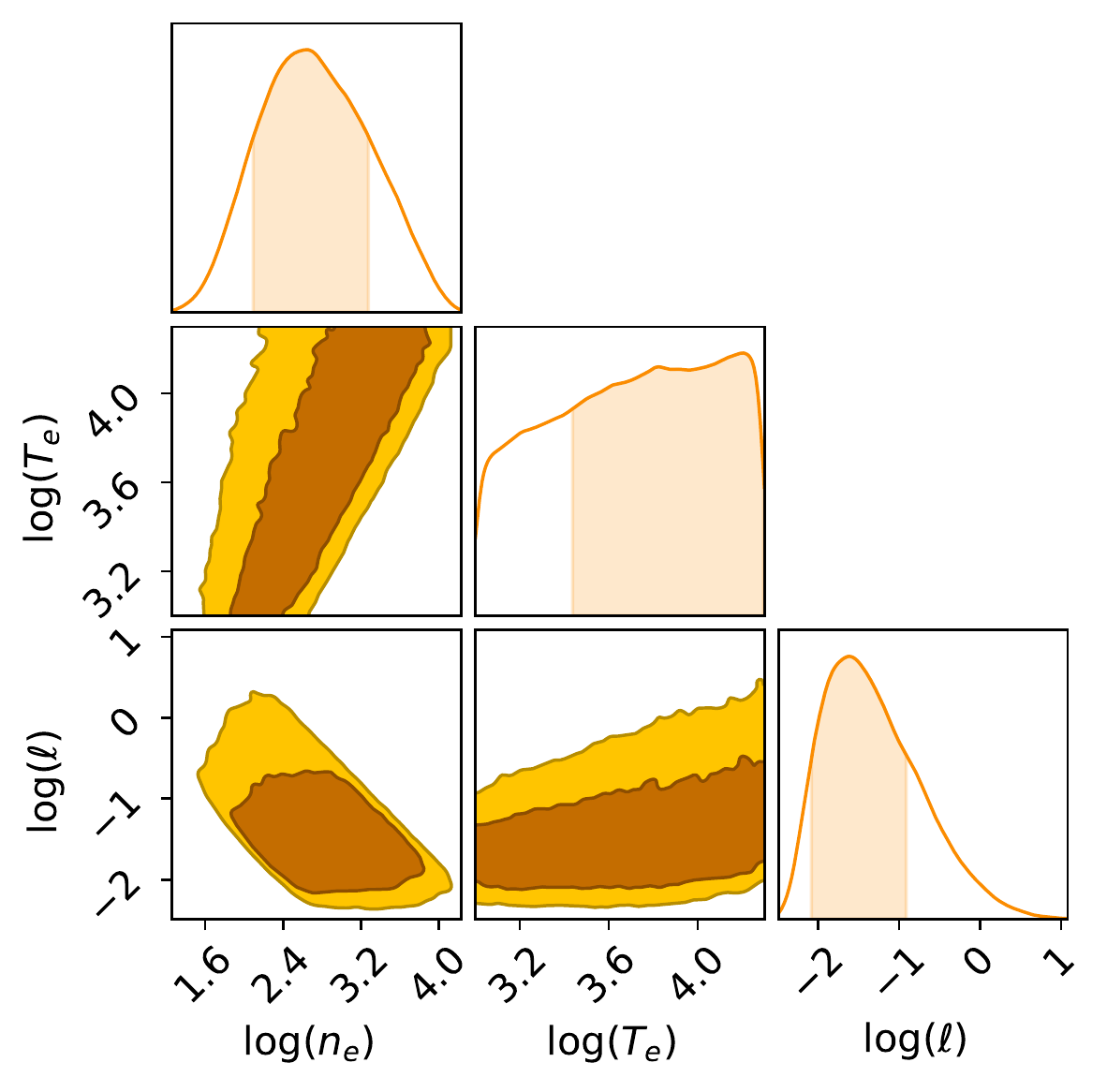}
    \caption{Corner plot showing the Total component constraints for the electron densities $n_e$ in units of \cmc{}, electron temperatures $T_e$ in units of K, and pathlength $\ell$ in units of pc. Contours are drawn at 68\% and 95\% intervals. The histograms show the marginalized distributions and the shaded region marks the 68\% credible intervals. }
    \label{fig:corner}
\end{figure}

The 2D and 1D marginal posterior distributions of the parameters of the Total component are plotted in Fig.~\ref{fig:corner}. They show that density is constrained by the models to within 0.6 dex with a maximum a posteriori value of $n_e = 500$~\cmc{} and 68.3\% credible interval of $130~\mathrm{cm^{-3}} \leq n_e \leq 2000$~\cmc{}, and that electron temperatures are not well constrained. Typical temperatures of photoionized gas have values of $T_e \approx 6000 - 10\,000$~K, with variations that are largely metallicity dependent \citep[e.g.,][]{Shaver1983, Wenger2019}. The RRL modeling shows that typical temperatures are slightly more likely than cooler temperatures. The maximum a posteriori value of the volume-averaged pathlength for the Total component is $\ell = 0.025$~pc and the 68.3\% credible interval is $7.9 \times 10^{-3}~\mathrm{pc} \leq \ell \leq 0.13$~pc. It is reasonable to infer that this ionized gas is not distributed in a very thin sheet of  $\ell \sim 0.025$~pc, but instead has a volume filling factor less than unity, and the emission arises from multiple, discrete clouds within the galaxy --- either as ionized clouds, interface layers of molecular clouds, \Hii{} regions (for further information see Sec.~\ref{ssec:sfr_ism}), etc.

\begin{figure*}
    \centering
    \includegraphics[width=\textwidth]{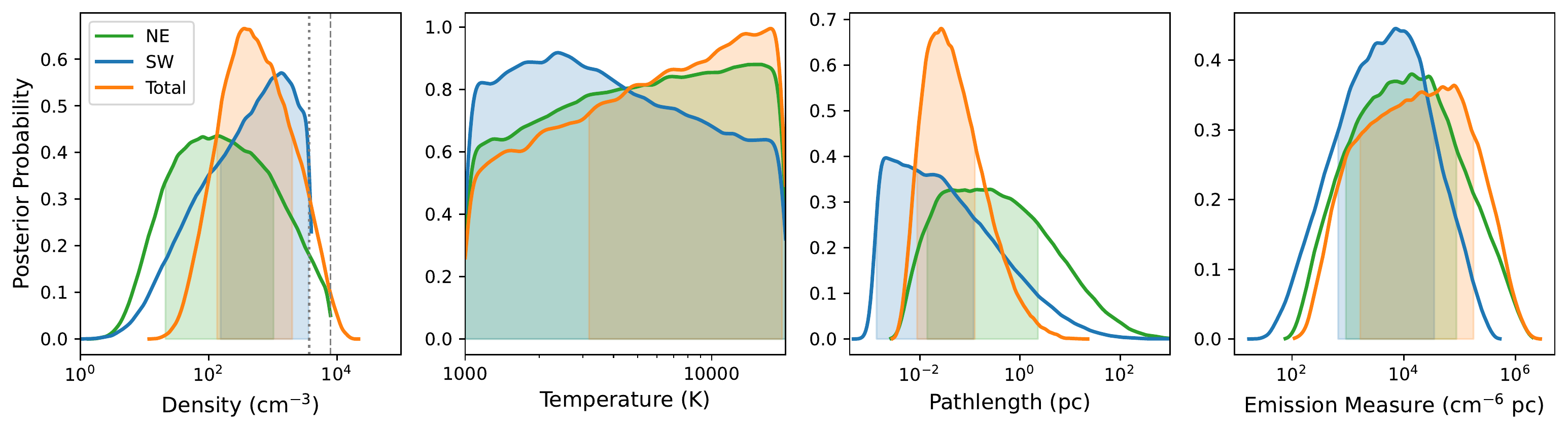}
    \caption{ Posterior probability distributions for the parameters constrained through the RRL observations: electron density $n_e$, electron temperature $T_e$, pathlength $\ell$, and emission measure $EM$, derived for the NE, SW, and Total components. The shaded regions represent the 68.3\% credible intervals. In the left-most panel, the gray dashed (dotted) line marks the $n_e < 7900$~\cmc{} ($<$3700~\cmc{}) constraint for the NE (SW) component obtained from the line width.}
    \label{fig:marg}
\end{figure*}

% table: physical conditions

\begin{deluxetable}{lcCCC}
\tablecaption{Constraints on physical conditions. Temperature is not well constrained within the allowed range $\log(T_e) = [3,4.3]$ K. \label{tab:physconds} }
\tablewidth{0pt}
\tablehead{
\colhead{} %1
& \colhead{$v_{\mathrm{cen}}$} %2
& \colhead{$\log( n_e ) $} %3
& \colhead{$\log( \ell ) $} %4
& \colhead{$\log( EM ) $} %4
\\
\colhead{} %1
& \colhead{(\kms{})} %2
& \colhead{(\cmc{})} %3
& \colhead{(pc)} %4
& \colhead{(cm$^{-6}$ pc)} %4
}
\startdata
NE   & -120     & 2.0_{-0.7}^{+1.0}      &  -0.7_{-1.1}^{+1.1}  & 4.1_{-1.2}^{+0.8} \\
SW   & 8        & 3.2_{-1.0}^{+0.4}      & -2.7_{-0.2}^{+1.8}   & 3.9_{-1.0}^{+0.7} \\
Total & (-230--55)  & 2.7_{-0.6}^{+0.6}  & -1.6_{-0.5}^{+0.7}   & 4.9_{-1.7}^{+0.4} \\
\enddata
%\tablecomments{}
\end{deluxetable}

In Fig.~\ref{fig:marg}, we show the marginal posterior distributions for all three components and each of the parameters, electron density ($n_e$), electron temperature ($T_e$), pathlength ($\ell$), and emission measure ($EM$). The parameter constraints are listed in Table~\ref{tab:physconds}, with the maximum a posteriori values and 68.3\% credible intervals. We do not list the temperature in Table~\ref{tab:physconds} because it is not well constrained by these measurements. We take 100 RRL models at random with properties that fall within the constraints of the 68.3\% credible intervals and plot the SLEDs of these models in Fig.~\ref{fig:SLED}, in order to demonstrate the variation in the predicted integrated optical depth for different models.

%%%%%%%%%%%%%%%%

\subsection{Mass of ionized gas}
\label{ssec:mass_ion}

Using the most likely values of the physical quantities in Table~\ref{tab:physconds}, we estimate the mass and mass per unit area of ionized gas in each component. These mass estimates are meant for qualitative comparisons as order of magnitude indications and do warrant caution. We assume the volume of gas is effectively described by a cylinder, as a circular region on the plane of the sky (core size) and with a distance into the plane equal to the path length. We also assume that the radio continuum emission is dominated by the NE and SW components, as these have been shown to account for $\sim99$\% of the emission at 1.4~GHz \citep{Koopmans2005, Combes2021}. To calculate the mass of ionized gas, we expect
\begin{eqnarray}
    M_{\mathrm{ion}} && \approx 1.36 m_{\mathrm{H}} \cdot n_e  \cdot (\pi r_{c}^2) \ell \nonumber \\
    M_{\mathrm{ion}} && \approx 0.11 \, \mathrm{M}_{\odot} 
        \left( \frac{n_e}{ \mathrm{cm}^{-3} } \right) 
        \left( \frac{r_c}{ \mathrm{pc} } \right)^2
        \left( \frac{\ell}{ \mathrm{pc} } \right).
\label{eq:m_ion}
\end{eqnarray}
where $m_{\rm H}$ is the hydrogen atom mass and $r_c$ is the radius of the radio continuum core. \cite{Guirado1999} constrained the source size of the SW component to follow $\propto \nu^{-2.0}$ resulting in a size of 0.1\asec{} at 1~GHz, which corresponds to the value we set of $r_c = 786$~pc at $z=0.89$. For the NE component we adopt the separation of 0.05\asec{} between the two brightest emission peaks as the size, $ r_c = 393$~pc. We do not include uncertainties for $r_c$ when computing the mass; however, we note that their errors are small in comparison to the large range uncertainty of the credible intervals. Using the posterior distributions of $n_e$ and $\ell$, we estimate a total mass for the NE component of $M_{\mathrm{ion}} \approx 10^{6.0_{-0.7}^{+0.6}}$~\Msun{} and for the SW component, $M_{\mathrm{ion}} \approx 10^{5.5_{-0.4}^{+0.8}}$~\Msun{}. If we assume that the area of the Total component is a summation of the areas intercepted by the NE and SW lines-of-sight, the estimated ionized mass is $M_{\mathrm{ion}} \approx 10^{6.4_{-0.5}^{+0.7}}$~\Msun{}.

It is informative to also calculate the gas mass per unit area
\begin{equation}
    \Sigma_{\mathrm{ion}} \approx 0.033 \, \mathrm{M}_{\odot} \,  \mathrm{pc}^{-2}
    \left( \frac{n_e}{ \mathrm{cm}^{-3} } \right) 
    \left( \frac{\ell}{ \mathrm{pc} } \right).
\end{equation}
For the NE, SW, and Total components, we calculate $\Sigma_{\mathrm{ion}}$ to be $10^{0.3_{-0.7}^{+0.6}}$~\Msunpc{}, 10$^{-0.8_{-0.4}^{+0.8}}$~\Msunpc{}, and $10^{0.0_{-0.5}^{+0.7}}$~\Msunpc{}, respectively. 

Using the surface densities of the NE, SW and Total components, we calculate an estimate for the total ionized gas mass of the galaxy within the Einstein ring of $R_{\rm g} \sim 5.3$~kpc in units of \Msun{} as $\log( M_{\rm ion,g} ) \approx 8.2_{-0.7}^{+0.6}$,  $7.1_{-0.4}^{+0.8}$, and $7.9_{-0.5}^{+0.7}$, respectively. While the mass estimated from the SW is more equivalent to a true lower limit of the total ionized gas mass, the estimates from the NE and Total components are almost certainly lower limits as well since we do not trace the bulk of the ionized gas mass of the galaxy contained in the Warm Ionized Medium \citep[WIM; see][]{Tielens2005}.

%%%%%%%%%%%%%%%%

\subsection{Ionizing photon flux}
\label{ssec:qnot_area}

We use the ionized gas emission measure to infer the ionizing photon flux. The ionizing photon rate, $Q_o$, per unit area is,
\begin{eqnarray}
    \frac{ Q_o }{\mathrm{area}} && \approx EM \cdot \alpha_B \nonumber \\ 
    \frac{ Q_o }{\mathrm{area}} && \approx 7.6 \times 10^{45} \,\mathrm{photons \, s^{-1} \, pc^{-2}}  \frac{EM}{10^3 \, \mathrm{cm^{-6} \, pc} } 
\end{eqnarray}
where $\alpha_B$ is the case B recombination coefficient. Using the posterior distributions of the emission measure, we calculate ionizing photon fluxes in units of \QAuni{} of $\log{(Q_o / \mathrm{area})} = 47.0_{-1.2}^{+0.8}$, $46.8_{-1.0}^{+0.7}$, and $47.8_{-1.7}^{+0.4}$ for the NE, SW, and Total components, respectively. These values are about an order of magnitude or more higher than the ionizing photon flux in the disk of the Milky Way \citep[][and references therein]{Kado-Fong2020}. 

While the gas mass estimates (Sec.~\ref{ssec:mass_ion}) are likely lower limits, the ionized photon fluxes are closer to realistic values, since they are dominated by the large emission measures that we are sensitive to.

%%%%%%%%%%%%%%%%%%%%%%%%%%%%%%%%%%%%%%%%

\section{Discussion} 
\label{sec:discuss}

%%%%%%%%%%%%

\subsection{Star Formation Rate and ISM Properties}
\label{ssec:sfr_ism}

The posterior distributions for the emission measure and thus directly the ionizing photon flux are constrained with fairly similar 68.3\% credible intervals. To estimate a star formation rate (SFR) of the galaxy, we adopt the log-average $EM$ within the credible intervals of the three components, $EM \approx 10^{4.0 \pm 0.8}$~\emuni{}, within a typical error of 0.8 dex, and therefore we adopt an ionizing photon flux is $Q_o / {\rm area} \approx 10^{46.9 \pm 0.8}$~\QAuni{}.
We estimate the total ionizing photon rate for the galaxy within $R_g = 5.3$~kpc as $Q_o = \pi R_g^2 \cdot 10^{46.9 \pm 0.8} \approx 10^{54.8 \pm 0.8}$~photons~s$^{-1}$. A Starburst99 model \citep{Leitherer1999} of continuous star formation establishes the relation with the ionizing photon rate (which levels off after $\sim$50~Myr) of $\mathrm{SFR} = 1~\mathrm{M_{\odot} \, yr^{-1}} \, (Q_o / 1.4 \times 10^{53} \, \mathrm{photons \, s^{-1}})$. Using this relation, an estimated SFR for the $z=0.89$ galaxy is $\mathrm{SFR} \approx 10^{1.7 \pm 0.8}$~\Msunyr{} and the SFR per unit area is $\Sigma_{\mathrm{SFR}} \approx 10^{-0.2 \pm 0.8}$~\Msunyr{}~kpc$^{-2}$. A galaxy on the main sequence at $z=0.89$ with SFR~$\sim$50~\Msunyr{} has a typical stellar mass of $\sim10^{11}$~\Msun{} \citep[e.g.,][]{Schreiber2015}. Current lensing models estimate the total mass within the Einstein ring as $M_E \approx 4 \times 10^{11}$~\Msun{} \citep{Muller2020b} and therefore a stellar mass of $\sim 8 \times 10^{10}$~\Msun{} given a typical mass to stellar light ratio of $\sim5$ \citep{Treu2004}. This is likely a main-sequence galaxy.

In Sec.~\ref{ssec:mass_ion} we estimated the ionized gas mass per unit area of the NE component to be $\Sigma_{\rm ion} \sim 2.1$~\Msunpc{}. For comparison, the \Hi{} column density is estimated to be $N_{\mathrm{H\,I}} \approx 5 \times 10^{21}$~cm$^{-2}$ assuming half of the continuum flux comes from the NE component \citep{Chengalur1999, Combes2021}. Assuming the average particle mass is $\approx 1.36m_{\rm H}$, the atomic gas mass per unit area is $\Sigma_{\mathrm{H\,I}} \approx 50$~\Msunpc{}. We use the OH column density of $N_{\rm OH} \approx 1.5 \times 10^{15}$~cm$^{-2}$ \citep{Gupta2021a} and assume an abundance of $10^{-7}$ \citep{Balashev2021} given that this line-of-sight has properties of a diffuse cloud \citep{Muller2011} in order to estimate a molecular gas column density of $N_{\rm H_2} \sim 1.5 \times 10^{22}$~cm$^{-2}$. This H$_2$ column density is higher than the $N_{\rm H_2} \sim 10^{21}$~cm$^{-2}$ derived from H$_2$O absorption, with a smaller continuum cross section at higher frequencies \citep{Muller2014a}. Assuming the average particle mass is $\approx 2m_{\rm H}$, the molecular gas mass per unit area is $\Sigma_{\rm H_2} \approx 240$~\Msunpc{}. 

With these estimates, the neutral gas mass per unit area is $\Sigma_{\rm HI + H_2} \sim 290$~\Msunpc{}.
The ionized gas mass detected in RRLs of the NE component is a small fraction ($\sim$0.7\%) of the total gas mass. 
The Kennicutt-Schmidt law \citep{Kennicutt1998a} long establishes a direct correlation between the surface densities of the neutral gas mass $\Sigma_{\rm HI + H_2}$ and the star formation rate $\Sigma_{\rm SFR}$ in galaxies on spatial scales 300--500 pc or more \citep{Schruba2010, Kruijssen2014a}. It suggests that a region in a galaxy with $\Sigma_{\rm SFR} \sim 0.6$~\Msunyr{}~kpc$^{-2}$ has a neutral gas mass per unit area of $\Sigma_{\rm HI + H_2} \sim 270$~\Msunpc{} \citep[e.g.,][]{Kennicutt2021}. Our measured neutral gas mass agrees to within 10\% of the expected value, although our measured SFR surface density has a large uncertainty.

Although the estimated densities of the NE, SW and Total components are consistent within the errors, the most likely density of the SW component, $n_e \approx 1600$~\cmc{}, is typical of young, compact \Hii{} regions. If we assume the \Hii{} regions are $r_{\rm{H\,II}} \approx 2$~pc in size, then the covering fraction through this cross section of the galaxy is $f_c \approx \ell_{\rm SW}/ (4/3 \cdot r_{\rm{H\,II}}) \approx 7.5 \times 10^{-4}$. The total number of \Hii{} regions, $N_{\rm{H\,II}}$, is calculated from the surface density estimates, $\pi r_C^2 = N_{\rm{H\,II}} / f_c \cdot (\pi\, r_{\rm{H\,II}}^2) $, which computes to $N_{\rm{H\,II}} \sim 116$, where we recall from Sec.~\ref{ssec:mass_ion} that $r_C = 786$~pc for the SW component. Equivalently, assuming all \Hii{} regions have $n_e \approx 1600$~\cmc{} and $r_{\rm{H\,II}} \approx 2$~pc, the ionized gas mass per \Hii{} region is estimated to be $M_{\rm H\,II} = 1860$~\Msun{}, and from the total ionized mass of the SW component, $M_{\rm SW} = N_{\rm H\,II} M_{\rm H\,II}$, also results in $N_{\rm H\,II} \approx (2.1 \times 10^5 \, \rm{M}_{\odot}) / (1860 \,  \rm{M}_{\odot}) \sim 113$. However, this calculation warrants caution because the number of \Hii{} regions changes dramatically depending on the assumed size. We set $r_{\rm H\,II} \approx 2$~pc because it is the typical size of young, massive star clusters \citep{Ryon2017} that are expected to dominate star formation in galaxies of this epoch, i.e., with high gas surface densities and star formation rates \citep{Kruijssen2012}.

%%%%%%%%%%%%

\subsection{The $n_e - \Sigma_{\rm SFR}$ relation }
\label{ssec:ne_Sigma}

The star formation rate per unit area is shown to be correlated with the electron density of ionized gas in the region \citep{Shimakawa2015, Herrera-Camus2016, Jiang2019}. Assuming \Hii{} regions thermalize with the ISM \citep{Gutierrez2010}, i.e., take on a pressure balance with others phases, then the volume-average density of the ionized gas (along with fairly consistent ionized gas temperatures) indicates the thermal pressure of the medium \citep{Jiang2019, Barnes2021}. This results in a $P-\Sigma_{\mathrm{SFR}}$ relation that serves as an important test-bed for pressure-regulated, feedback-modulated star formation \citep[e.g.,][]{Kim2013, Ostriker2022}. Using doublet ratios of [SII] and [OII] in the optical \citep{Kewley2019}, \cite{Shimakawa2015} and \cite{Jiang2019} find a relation of $\Sigma_{\mathrm{SFR}} \propto n_e^{1.7 \pm 0.3}$ in a sample of $z \sim 1-3$ starburst galaxies. In the far IR, the ratio of [NII] fine structure lines from a sample of nearby normal galaxies and (ultra) luminous infrared galaxies ((U)LIRGs; $L_{\mathrm{IR}} \ge 10^{11}$~\Lsun{}) establish $\Sigma_{\mathrm{SFR}} \propto n_e^{1.5}$ \citep{Herrera-Camus2016}.

For $\Sigma_{\mathrm{SFR}} \sim 0.6$~\Msunyr{}~kpc$^{-2}$, the electron density predicted by the \cite{Shimakawa2015} relation is $\sim 110$~\cmc{} and by the \cite{Herrera-Camus2016} relation is $\sim 200$~\cmc{}. These density estimates agree well with the NE component and lie on the cusp of the 68\% credible intervals of the SW and Total components.
If the RRL emission is tracing the thermal properties of the diffuse medium in the galaxy's disk, higher pressures and densities are typically found at smaller galactic radii, i.e., the SW component, than at larger galactic radii, i.e., the NE component.
For example, \cite{Gutierrez2010} measured the electron density to increase towards smaller galactic radii, $r$, as $\left< n_e \right> = \left< n_e \right>_o \exp{(-r / R_g)}$, where $R_g$ is the scale length of the disk at which star-formation and density drops off and $\left< n_e \right>_o$ is inner most density.  With $R_g = 5.3$~kpc and $\left< n_e \right> = 100$~\cmc{} of the NE component, at $r=2.4$~kpc the expected density is $n_e \sim 160$~\cmc{}. While the contrast in density at the two radii is not as extreme as the best fit densities of the components indicate, the general trend is consistent and a density of 160~\cmc{} does fall within the 68\% credible interval of the SW component. The smaller pathlength in comparison to the NE component might then indicate a smaller covering fraction of the overall larger cross section of this line-of-sight, i.e. only within or close to the spiral arm.

%%%%%%%%%%%%

\subsection{Radio Continuum SED}
\label{ssec:radioSED}

The radio continuum emission from \trg{} is complex. The lensing is achromatic and contains a small-scale core-jet structure with regions of different spectral indices and opacities. In addition, the source is variable on hourly to yearly timescales \citep{PrameshRao1988, vanOmmen1995, Lovell1996, Lovell1998b, Garrett1997, Jin2003, Marti-Vidal2013, Allison2017}, which makes it difficult to compare observations and model the radio continuum emission \citep{Muller2020b}. 

Ionized gas that is detectable in RRLs and has a large covering fraction would also emit free-free emission and when it becomes optically thick, would absorb any background radio continuum. A reliable model and knowledge of the spatially-resolved radio SED could independently constrain the physical conditions of the gas through free-free absorption. The frequency at which radio emission becomes optically thick to free-free absorption is defined by $\tau_\nu = 6.67 \times 10^{-2} \, EM \, T_e^{-1.323} (\nu / \mathrm{GHz})^{-2.118}$ \citep[e.g.,][]{Emig2022}. 

\begin{figure}
    \centering
    \includegraphics[width=0.47\textwidth]{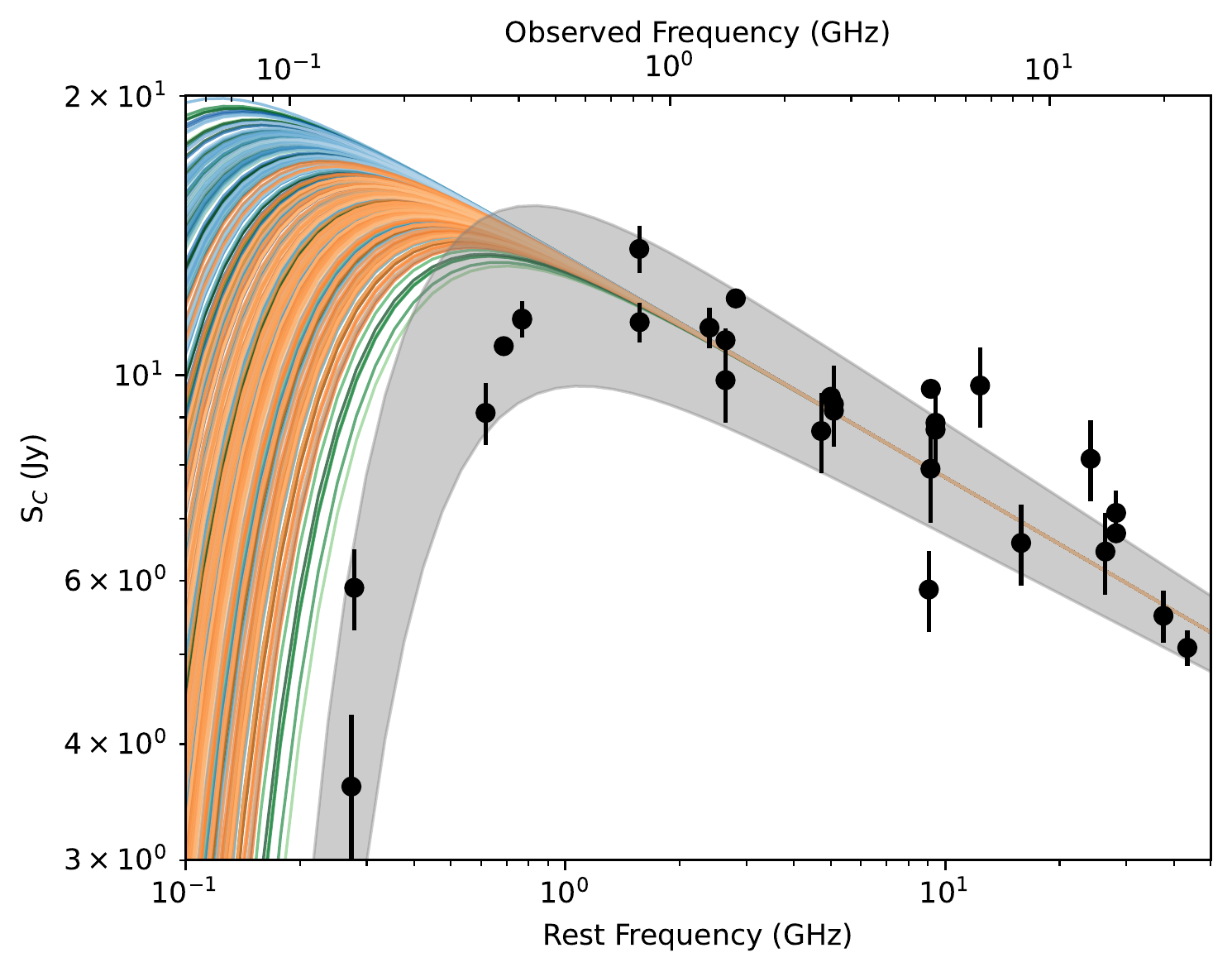}
    \caption{ Compilation of continuum flux density measurements from the bright and highly variable (factors of $\sim$1.5 on weeks and years timescales) \trg{}. The gray shaded region encompasses the 1$\sigma$ confidence region of a best-fitting power-law that is attenuated by free-free absorption. We also include the continuum SEDS (with colors corresponding to the components on the left hand plot) from a fixed power-law and that is attenuated by ionized gas with properties constrained by the RRL models. We assume a redshift $z=0.89$ for the conversion between observed and rest frequencies. }
    \label{fig:SED}
\end{figure}

We collected radio continuum measurements of \trg{} at $\nu \lesssim 22$~GHz \citep[using the NASA/IPAC Extragalactic Database (NED) and][]{PrameshRao1988, Henkel2008, Intema2017} and plot these in Fig.~\ref{fig:SED}. We fit the SED of \trg{} as a power-law index with an external screen of free-free absorption, using the functional form $S_{\nu} = S_o (\frac{\nu}{\nu_o})^{\alpha} \exp (- \tau_{\nu} ) $ with $\tau_{\nu} = \tau_o (\frac{\nu}{\nu_o})^{-2.118}$. Setting $\nu_o = 40$~GHz with respect to observed frequencies, the best fit parameters with 1$\sigma$ uncertainties are $S_o = 4.8 \pm 0.4$~Jy, $\alpha = -0.24 \pm 0.03$ and $\tau_o = (1.1 \pm 0.2) \times 10^{-5}$.

In Fig.~\ref{fig:SED}, we also plot how ionized gas that has physical properties constrained by the RRL models --- using the same selection of models presented in Fig.~\ref{fig:SLED} --- would attenuate a power-law continuum SED (normalized by $S_o$ and $\alpha$ of our fit). The emission from the RRLs we detect at $z=0.89$ would result in free-free absorption at lower frequencies than is observed in \trg{}.
For the $z=0.89$ galaxy to cause the free-free absorption, volume-average pathlengths a factor of 5 larger are needed. A smaller filling factor and larger pathlength intercepting the radio emission is not unreasonable.

It would still be possible for ionized gas in the environment of the blazar at $z=2.5$ to create the absorption in the radio SED. Even though we do not detect hydrogen RRL emission at $z=2.5$, ionized gas with different properties, for example higher densities, could be present and still be consistent with the RRL constraints. 
We also note that ionized gas in the Milky Way with $EM$s that result in a turnover at the observed frequencies (especially, on $<1$\asec{} scales) would be observable in hydrogen RRL emission at $z=0$, and we do not detect any RRL emission from the Milky Way.

%%%%%%%%%%%%%%%%%%%%%%%%%%%%%%%%%%%%%%%%

\section{Conclusions}
\label{sec:conclude}

We used MALS observations to detect RRL emission in the spectrum of the radio blazar \trg{}. The RRL emission is observed at $z=0.89$ from a galaxy that lies along the line of sight and strongly lenses \trg{}. This is the second detection of RRLs outside of the local universe (i.e., at $z \geq 0.076$) and the first clearly associated with hydrogen \citep[e.g.,][]{Emig2019}. We detect H144$\alpha$ by stacking 17 RRLs covered by the L band (856--1712 MHz) with a S/N of 21 (see Fig.~\ref{fig:spec_z0p89_L}), and we detect H163$\alpha$ by stacking 27 lines in the UHF band (544--1088 MHz) with a S/N of 14 (see Fig.~\ref{fig:spec_z0p89_UHF}). Emission from the H144$\alpha$ line is consistent over two separate observations, when comparing the spectra in parallel hand polarizations, and is robust against additional spectral stacking verification methods. Like the \Hi{} 21~cm and OH 18~cm absorption spectra (see Fig.~\ref{fig:compare_tracers}), the H144$\alpha$ and H163$\alpha$ emission profiles span $\sim$250~\kms{} in velocity and are dominated by two velocity components associated with two physically distinct regions of the galaxy, the NE and SW lines of sight. We do not detect RRL emission in either band intrinsic to \trg{} ($z=2.5$), from the $z=0.19$ absorption system along this line of sight, or from the Milky Way (see Fig.~\ref{fig:spec_nondetect}). 

Hydrogen RRL emission typically arises from fully ionized gas and only stimulated emission is observable outside of the local universe. The maser-like properties of stimulated emission enable the RRL SLED to constrain the density and pathlength of the ionized gas (see Table~\ref{tab:physconds} and Fig.~\ref{fig:marg}). Considering the total integrated line intensity, referred to as the Total component, we used a Bayesian analysis to constrain the electron density of the gas $\log( n_e ) = 2.6 \pm 0.6$~\cmc{} and a volume-averaged pathlength of $\log(\ell) = -1.6_{-0.5}^{+0.7}$~pc, which likely has a non-unity filling factor. Analyzed separately, the NE line-of-sight appears to harbor less dense gas with $\log( n_e ) = 2.0_{-0.7}^{+1.9}$~\cmc{} and $\log( \ell ) = -0.7 \pm 1.1 $~pc, and the SW line-of-sight appears to intercept dense gas that is more typical of \Hii{} regions with $\log( n_e ) = 3.2_{-1.0}^{+0.4}$~\cmc{} and $\log( \ell ) = -2.7_{-0.2}^{+1.8}$~pc. These scenarios are consistent with the NE line-of-sight passing through diffuse clouds at a larger galactic radius, and the SW component directly intercepting a spiral arm, as has previously been determined.

The RRL components measure an ionizing photon flux of $Q_o / \mathrm{area} \approx 10^{46 \pm 0.8}$~\QAuni{} and star formation rate surface density of $\Sigma_{\rm SFR} \sim 10^{-0.2 \pm 0.8}$~\Msunyr{}~kpc$^{-2}$. Taken over the $z=0.89$ galaxy within $R_g \sim 5.3$~kpc, the ionizing photon rate of $Q_o \sim 10^{54.8}$~photons~s$^{-1}$ yields an average star formation rate of SFR~$\sim 50$~\Msunyr{}. Despite the plethora of molecular species observed, the ionized gas content and SFR have not been previously measured for this source, largely due to the highly reddened nature of \trg{} at optical and NIR wavelengths. In comparing the SFR and the galaxy's mass (from lensing), the $z=0.89$ system is likely on the main sequence.

The ionized gas mass per unit area of the diffuse NE component as measured by the RRL emission is $\Sigma_{\rm ion} \approx 2.1$~\Msunpc{}, in comparison with gas masses of $\Sigma_{\mathrm{H\,I}} \approx 50$~\Msunpc{} and $\Sigma_{\rm H_2} \approx 240$~\Msunpc{} (via OH) estimated from only the MALS observations. Given our estimated SFR, the \Hi{}+H$_2$ gas mass surface density is close to the gas content predicted by the Kennicutt-Schmidt law. Our measured electron densities also match reasonably well with the $n_e - \Sigma_{\mathrm{SFR}}$ relation determined from optical and FIR line ratios.

% Closing
\trg{} is the first source investigated with MALS, and the detection of RRLs in the source is promising for the remaining $\sim$500 targets of the survey. With the first hydrogen RRL detection that breaks the redshift-barrier, we show that this tracer can be an important tool for investigating (a) the electron density (thermal pressure) of ionized gas in the ISM of galaxies (and the $n_e - \Sigma_{\rm SFR}$ relation) and (b) the SED of AGN, thus eventually AGN evolution. We have also demonstrated the unique science that can be achieved through \Hi{} 21~cm, OH 18~cm, and RRL measurements that are simultaneously observed in the MALS survey. The ionized gas properties in the $z=0.89$ galaxy will be substantially improved through RRL observations at higher and lower radio frequencies and at higher ($<$1\asec{}) spatial resolutions which can separate the two main (velocity) components of emission. The new science afforded by high-redshift RRL studies is accessible with on-going wide-bandwidth spectral line surveys and will be explored in unprecedented capacities with future facilities such as the next generation Very Large Array \citep[ngVLA;][]{Murphy2018} and the SKA \citep{Carilli2015}.

%%%%%%%%%%%%%%%%%%%%%%%%%%%%%%%%%%%%%%%%

\acknowledgments

The authors acknowledge and appreciate the efforts and input of the anonymous reviewer of this article. The authors thank Peter Shaver for comments on the article and for the inspiration and motivation to carry through with this research.

SAB was supported by RSF grant 18-12-00301. The MeerKAT telescope is operated by the South African Radio Astronomy Observatory, which is a facility of the National Research Foundation, an agency of the Department of Science and Innovation. 
The MeerKAT data were processed using the MALS computing facility at IUCAA (https://mals.iucaa.in/releases)
The National Radio Astronomy Observatory is a facility of the National Science Foundation operated under cooperative agreement by Associated Universities, Inc. This research has made use of the NASA/IPAC Extragalactic Database (NED), which is funded by the National Aeronautics and Space Administration and operated by the California Institute of Technology. 

%% To help institutions obtain information on the effectiveness of their 
%% telescopes the AAS Journals has created a group of keywords for telescope 
%% facilities.
%
%% Following the acknowledgments section, use the following syntax and the
%% \facility{} or \facilities{} macros to list the keywords of facilities used 
%% in the research for the paper.  Each keyword is check against the master 
%% list during copy editing.  Individual instruments can be provided in 
%% parentheses, after the keyword, but they are not verified.

\facilities{MeerKAT}

%% Similar to \facility{}, there is the optional \software command to allow 
%% authors a place to specify which programs were used during the creation of 
%% the manuscript. Authors should list each code and include either a
%% citation or url to the code inside ()s when available.

\software{
ARTIP \citep{Gupta2021a},
Astropy \citep{Astropy2018, Astropy2022},
CASA \citep{McMullin2007, CASATeam2022},
ChainConsumer \citep{Hinton2016},
CRRLpy \citep{Salas2016},
emcee \citep{Foreman-Mackey2013},
Matplotlib \citep{Hunter2007},
and NumPy \citep{Harris2020}
}

%%%%%%%%%%%%%%%%%%%%%%%%%%%%%%%%%%%%%%%%

\bibliography{Emig2022_pks1830}{}
\bibliographystyle{aasjournal}

%%%%%%%%%%%%%%%%%%%%%%%%%%%%%%%%%%%%%%%%

%% Appendix material should be preceded with a single \appendix command.
%% There should be a \section command for each appendix. Mark appendix
%% subsections with the same markup you use in the main body of the paper.

%\appendix

%\section{Millimeter wavelength emission from free-free continuum and recombination lines}
%\label{ap:derivation}

%% This command is needed to show the entire author+affiliation list when
%% the collaboration and author truncation commands are used.  It has to
%% go at the end of the manuscript.
%\allauthors

%% Include this line if you are using the \added, \replaced, \deleted
%% commands to see a summary list of all changes at the end of the article.
%\listofchanges

\end{document}